\documentclass[twocolumn]{article}

\RequirePackage{amsthm,amsmath,amsfonts,amssymb}
\RequirePackage[authoryear]{natbib}
\RequirePackage{graphicx}

\usepackage{amssymb, amsfonts, xcolor, bm, float, bbm, multirow}
\usepackage{algorithm} 
\usepackage{algpseudocode} 
\usepackage{diagbox} 
\usepackage{hyperref} 
\usepackage{booktabs} 
\DeclareMathOperator{\train}{train}
\DeclareMathOperator{\test}{test}
\newcommand{\vgood}[1]{\colorbox{blue!25}{#1}}
\newcommand{\best}[1]{\colorbox{blue!50}{\textbf{#1}}}
\newlength{\mytextwidth} 
\usepackage{enumitem}
\setlist{itemsep=2pt, topsep=2pt}

\usepackage[a4paper,twocolumn,margin=14mm,textwidth=528pt]{geometry}

\usepackage{etoolbox}
\AtBeginEnvironment{tabular}{\small}

\begin{document}

\title{A Comparison of Kernels for ABC-SMC}
\author{\large Dennis Prangle$^{1}$\thanks{dennis.prangle@bristol.ac.uk}, \quad
        Cecilia Viscardi$^{2}$\thanks{cviscardi@unisa.it}, \quad
        Sammy Ragy$^{1}$\thanks{sammyragy@gmail.com} \\ 
        $^{1}$University of Bristol \ 
        $^{2}$University of Salerno \ 
        }
\date{}
\maketitle

\begin{abstract}
    A popular method for likelihood-free inference is
    approximate Bayesian computation sequential Monte Carlo (ABC-SMC) algorithms.
    These approximate the posterior using a population of particles,
    which are updated using Markov kernels.
    Several such kernels have been proposed.
    In this paper we review these, highlighting some less well known choices,
    and proposing some novel options.
    Further, we conduct an extensive empirical comparison of kernel choices.
    Our results suggest using a one-hit kernel with a mixture proposal as a default choice.
\end{abstract}

\section{Introduction} \label{sec:intro}

\textbf{Likelihood-free inference} (LFI) or \textbf{simulation-based inference} (SBI)
refers to the task of statistical inference
when numerical evaluation of the likelihood is infeasibly expensive
but it is possible to repeatedly simulate datasets $y$ from the model given parameter values $\theta$.
\textbf{Approximate Bayesian Computation} (ABC) is a popular and well-studied class of Monte Carlo algorithms for this problem
\citep{marin2012approximate, sisson2018abc, beaumont2019approximate}.
ABC algorithms simulate data under many parameter values and
assign weights based on a distance between the simulated and observed data.
A simple case, which we use throughout, is to let the weight equal one when the distance is below a threshold $\epsilon$,
and zero otherwise.
The weighted sample of parameter values is used to approximate the posterior distribution.
The accuracy of the approximation depends on $\epsilon$.

Many ABC sampling schemes have been proposed.
We focus on ABC-SMC, a popular \textbf{sequential Monte Carlo} version of ABC
where the set of parameters used for simulation are updated sequentially,
so that the algorithm focuses on performing simulations for promising parameter values.
The algorithm iteratively reduces $\epsilon$, improving the accuracy of the posterior approximation.

At each iteration, ABC-SMC maintains a set of \textbf{particles}, in our case $(\theta, y)$ pairs, and probabilistically updates each particle to a new one.
A crucial tuning choice in ABC-SMC is which \textbf{Markov kernel} to use for the updating step.
We consider \textbf{ABC-MCMC} kernels,
which have the property of leaving an appropriate distribution (detailed later) for $(\theta, y)$ invariant.
The $\theta$ marginal of this is an approximation to the posterior.

The aim of this paper is to review and empirically compare kernels used in ABC-SMC.
In doing so, we propose some novel kernel choices, and highlight little known options from the literature.

We consider two components of a kernel.
The first is a \textbf{proposal kernel} which proposes updates for the parameters $\theta$.
We will consider the case where this can be expressed using conditional densities $q(\theta' | \theta)$.
Secondly, a \textbf{kernel family} maps a choice of $q$ to a transition kernel on $(\theta,y)$
which maintains invariance on the target distribution.

The simplest kernel family that we consider is \textbf{ABC Metropolis-Hastings} (ABC-MH).
This samples $\theta'$ from $q(\theta' | \theta)$ and data $y'$ from the model under $\theta'$.
Then it accepts and returns $(\theta',y')$ or rejects and returns $(\theta,y)$.
We also consider more complicated kernel families, including ``$r$-hit'' methods \citep{lee2012choice}.
These perform a random number of samples from $q$ and corresponding model simulations
before the final output is eventually returned.

Most papers use a Gaussian random walk proposal $q(\theta' | \theta)$,
with the covariance based on the particles of the previous ABC-SMC iteration.
(The exact choice of covariance varies: see Section \ref{sec:proposals}.)
In this paper we also investigate proposals using density estimation of the previous particles.
The potential benefit is to reduce time simulating data under poor parameter values.
Rather than random walk proposals, these are \textbf{independence proposals}:
the distribution for $\theta'$ under $q(\theta' | \theta)$ does not depend on $\theta$.
One independence proposal we consider is that of \cite{bernton2019approximate}
who use a Gaussian mixture distribution trained on the previous particles.

Our contributions are:
\begin{enumerate}
    \item A novel kernel family: a Metropolis-Hastings approach which is a simple alternative to $r$-hit kernels, for the case of independence proposals.
    \item A novel independence proposal, based on normalising flows.
    \item A review of existing kernel families and proposals for ABC-SMC and related issues.
    \item A large empirical comparison of kernel families and proposals on a range of models.
          We argue the results support a default choice of using the one-hit kernel family with a mixture proposal.
\end{enumerate}

In the remainder of the paper,
Section \ref{sec:abc} gives an overview of ABC and ABC-SMC.
Sections \ref{sec:kernel_families} and \ref{sec:proposals} describe kernel families and proposal kernels respectively.
Both combine a review of existing material and novel contributions.
Section \ref{sec:illustrations} illustrates the kernels on a simple example.
Section \ref{sec:experiments} reports our experiments,
and Section \ref{sec:conclusion} concludes with a discussion.
Our code is at \url{www.github.com/dennisprangle/abc_smc_kernel_comparison}.
All experiments use a virtual machine with 1 core and 2 Gb memory.

\subsection{Related work} \label{sec:related}

In addition to ABC-SMC,
another popular class of sequential ABC algorithms is ABC-PMC \citep{sisson2007sequential,beaumont2009adaptive,toni2009approximate}.
This can be viewed as iteratively performing an ABC version of importance sampling for a decreasing sequence of $\epsilon$ values.
Each proposal distribution for $\theta$ is generated adaptively using the output of the previous iteration.
A popular proposal in ABC-PMC -- from \citet{beaumont2009adaptive} --
is a version of kernel density estimation:
a $\theta$ value from the output is uniformly sampled and Gaussian noise is added.
The covariance of the noise is based on the particles of the previous iteration,
with theoretical justification based on the Kullback Leibler divergence between the proposal and ABC posterior for $\theta$.
\cite{filippi2013optimality} present variations on this analysis and proposal.
Recently, \cite{picchini2022guided} review more sophisticated ABC-PMC proposals.

ABC-SMC requires kernels which perform effective MCMC updates on $(\theta, y)$.
In contrast ABC-PMC requires proposals for $\theta$ which will perform effectively in ABC importance sampling.
Hence the overall requirements for good kernels are different.
Nonetheless, a common ABC-SMC proposal kernel for $\theta$
is based on the ABC-PMC $\theta$ proposal of \cite{beaumont2009adaptive}.
We review this later in Section \ref{sec:classicRW}.

\cite{everitt2021delayed} propose a ABC-MCMC kernel based on delayed acceptance for use in ABC-SMC.
This involves an initial accept/reject step based on a cheap surrogate likelihood.
On acceptance, a slower second stage is performed.
This simulates a dataset to make a final accept/reject decision.
We do not consider this approach in our paper,
as it is specialised to the case where a cheap surrogate is available.
But we expect it would be possible to create delayed acceptance versions of many of the kernels reviewed in our paper.

Recently, \cite{cao2024using} propose a ABC-SMC method with an adaptive early rejection MCMC kernel.
This method uses a Gaussian process to model the discrepancy between simulated and observed data, and early rejects particles using an acceptance ratio based on the modelled discrepancy.
It would be interesting to compare this to other ABC-MCMC kernels in future work.

\cite{arbel2021annealed} and \cite{matthews2022continual} consider a related problem to us in a likelihood-based SMC algorithm.
Rather than adapting the MCMC kernel, they adaptively learn a deterministic transport map to apply to the particles.
We build on parts of their methodology for our novel kernel family in Section \ref{sec:tabc}.

\section{Approximate Bayesian computation review} \label{sec:abc}

This section reviews ABC, focusing on the ABC-SMC algorithm.
For more general reviews see \citet{marin2012approximate}, \citet{sisson2018abc}, \citet{beaumont2019approximate}.
Section \ref{sec:abc_intro} introduces ABC and overviews different ABC algorithms.
Then Section \ref{sec:abc_smc} gives a more detailed description of the ABC-SMC algorithm.

\subsection{Introduction to ABC} \label{sec:abc_intro}

Consider the setting of Bayesian inference
with a prior density $\pi(\theta)$ on parameters $\theta$ and a likelihood $f(y | \theta)$.
We use $y_0$ to denote a specific observed dataset,
and $y$ to denote a generic dataset (often simulated data below).
The posterior density is then $p(\theta | y_0) \propto \pi(\theta) f(y_0 | \theta)$.
Throughout we assume $\theta \in \mathbb{R}^k$ and that densities for $\theta$ are with respect to Lebesgue measure.

Often the likelihood is intractable: numerical evaluation is impossible or impractically expensive.
Therefore standard likelihood-based inference is not possible.
In many situations, however, it is possible to simulate data from the model relatively cheaply.
This allows approximate inference by SBI/LFI methods, such as ABC.

The simplest ABC algorithm is \textbf{rejection ABC}.
Here parameter values are sampled from the prior
and each proposal $\theta$ is used to simulate a dataset $y$. A pair $(\theta, y)$ is accepted if $y$ is close to the observed data,
in the sense that $d(y,y_0) \leq \epsilon$ for some distance function $d$ and threshold $\epsilon$.
For simplicity we use Euclidean distance for $d$ throughout our paper, but many other choices are possible
\citep{bernton2019approximate, drovandi2022comparison}.

Accepted pairs are samples from
\begin{align}
    p_\epsilon(\theta, y | y_0) & \propto \pi(\theta) f(y | \theta) \mathbbm{1}[y \in B_\epsilon],
    \quad \text{where} \label{eq:ABC_jointposterior}                                               \\
    B_\epsilon                  & =\{ y | d(y,y_0) \leq \epsilon \} \label{eq:Beps}
\end{align}
is the set of $y$ values resulting in acceptance.
The marginal for $\theta$ is the \textbf{ABC posterior}
\begin{equation}
    \label{eq:ABC_posterior}
    p_\epsilon(\theta | y_0) \propto \pi(\theta) \int f(y | \theta) \mathbbm{1}[y \in B_\epsilon] dy.
\end{equation}
This is an approximation of the posterior density.
In the limit $\epsilon \rightarrow 0$ it converges to $p(\theta | y_0)$ (see Theorem 1 in the supplement of \citealp{prangle2017adapting}).

Rejection ABC is simple but inefficient,
as most time is typically spent performing simulations under parameter values with low ABC posterior density.
This motivates ABC algorithms which propose parameters more efficiently.

ABC-MCMC aims to improve efficiency by producing a Markov chain on $(\theta, y)$
with stationary density \eqref{eq:ABC_jointposterior}.
It uses a transition kernel to probabilistically update a pair $(\theta, y)$ to $(\theta', y')$.
Iterating the kernel produces a Markov chain which can be used for inference.
See \citet{sisson2011likelihood} for an overview.

The paper concentrates on ABC-SMC.
A large number of $(\theta, y)$ particles are sampled from the prior
and sequentially updated to target \eqref{eq:ABC_jointposterior}
with successively smaller $\epsilon$ values.
Updating the particles involves using ABC-MCMC kernels.
Section \ref{sec:abc_smc} gives a more detailed description of ABC-SMC.

We focus on the choice of ABC-MCMC kernel when used as a component of ABC-SMC.
Sections \ref{sec:kernel_families} and \ref{sec:proposals} review ABC-MCMC kernels
for this purpose,
including existing choices and some novel variations.

All ABC algorithms typically perform poorly unless $\dim y$ is small.
Otherwise, producing acceptances in a reasonable time
can require a large $\epsilon$ value which produces significant approximation error.
Hence $y$ is often replaced with lower dimensional \textbf{summary statistics} $s(y)$.
See \cite{prangle2018summary} and \cite{chen2023learning} for reviews.
As this issue is not the main focus of our paper,
our examples later use low $\dim y$ or some prespecified summary statistics.

\subsection{ABC-SMC} \label{sec:abc_smc}

ABC-SMC is a special case of a SMC (sequential Monte Carlo) sampler \citep{del2006sequential}.
Here we describe the algorithm and its properties.
We omit the theoretical justification, as in this paper it is unchanged from the existing literature.

The name ``ABC-SMC'' is sometimes also used to refer to
the ABC-PMC algorithms summarised in Section \ref{sec:related},
but we avoid this usage throughout our paper.

Algorithm \ref{alg:ABC-SMC} outlines a version of the ABC-SMC which is a specific instance of that proposed by \citet{del2012adaptive}.
Iteration $t$ involves applying a ABC-MCMC kernel to each particle.
This should be designed to produce stationary density \eqref{eq:ABC_jointposterior} for the current threshold $\epsilon^t$.
The main aim of our paper is to compare different choices of ABC-MCMC kernel here.

There is some freedom in how the steps in Algorithm \ref{alg:ABC-SMC} are implemented.
We generally follow \citet{bernton2019approximate}, who made several improvements.
These implementation details are described in Sections \ref{sec:termination}--\ref{sec:resampling}.
Then Section \ref{sec:asymptotic} summarises theoretical results on ABC-SMC output,
and Section \ref{sec:ABCSMCvariations} discusses some variations on Algorithm \ref{alg:ABC-SMC}.

\begin{algorithm}[htbp]
    \caption{ABC-SMC} \label{alg:ABC-SMC}
    \begin{algorithmic}[1]
        \Statex Input: numbers of particles $N$; choice of ABC-MCMC kernel
        \State Initialise: $\epsilon^0 = \infty$,
        $(\theta^0_i, y^0_i)_{1 \leq i \leq N} \sim \pi(\theta) f(y|\theta)$ (independent samples).
        \For{$t=1,2,3,\ldots$}
        \State Calculate $\epsilon^t$ (see Section \ref{sec:threshold}).
        \State Calculate weights $w^t_i = \mathbbm{1}[y^{t-1}_i \in B_{\epsilon_t}]$.
        \State Resample particles (see Section \ref{sec:resampling}) to produce
        $(\theta'^{t-1}_i, y'^{t-1}_i)_{1 \leq i \leq N}$.
        \For{$i=1,2,\ldots,N$}
        \State Apply ABC-MCMC kernel to $(\theta'^{t-1}_i, y'^{t-1}_i)$
        to output $(\theta^t_i, y^t_i)$.
        \EndFor
        \EndFor
    \end{algorithmic}
\end{algorithm}

\subsubsection{Termination} \label{sec:termination}

We terminate Algorithm \ref{alg:ABC-SMC} after a prespecified runtime -- one hour in our experiments later.
We return results from the final complete iteration.
That is, our output is $(\theta^{t-1}_i)_{1 \leq i \leq N}$,
given that iteration $t$ is being performed at termination.
The output are approximate samples from the ABC posterior \eqref{eq:ABC_posterior} under the threshold $\epsilon^{t-1}$.

\subsubsection{Updating $\epsilon$} \label{sec:threshold}

\citet{del2012adaptive} propose a rule to select $\epsilon^t$.
In our setting (i.e.~where particle weights are zero or one)
this reduces to picking $\epsilon^t$ so that the number of particles with non-zero weight equals $\omega N$,
given a tuning value $\omega \in (0,1)$.
However, this can result in poor particle diversity
if there are many identical copies of particles after resampling.

Instead, we follow \citet{bernton2019approximate} who aim to produce least $\omega N$ unique particles.
That is, there should be at least $\omega N$ different $(\theta, y)$ values after resampling.
A bisection algorithm is used to find $\epsilon^t$ which satisfies this criterion.
This requires repeatedly calculating the number of unique particles resulting from
steps 4 and 5 of Algorithm \ref{alg:ABC-SMC}.
To make this a deterministic calculation, the random draws used for the resampling step are fixed
(and reused when step 5 is reached after $\epsilon^t$ is chosen).

Throughout our numerical experiments, we use the tuning choice $\omega = 0.5$.

\subsubsection{Resampling} \label{sec:resampling}

Resampling is a standard step in SMC algorithms.
It replaces the current weighted particles with a new set of equally weighted particles.
The new particles are sampled from the current particles with probabilities proportional to their weights.
In Algorithm \ref{alg:ABC-SMC}, triples
$(\theta^{t-1}_i, y^{t-1}_i, w^t_i)_{1 \leq i \leq N}$ are updated to
$(\theta'^{t-1}_i, y'^{t-1}_i, 1)_{1 \leq i \leq N}$.
(That is, the new weights all equal 1 but are not recorded notationally in the algorithm.)

Sampling each new particle independently is known as multinomial resampling.
However, dependent resampling schemes are more efficient \citep{chopin2020introduction}.
Following \citet{bernton2019approximate}, we use systematic resampling \citep{carpenter1999improved}.
This is recommended as a default choice \citep{chopin2020introduction},
although more complex alternatives exist with improved theoretical properties \citep{gerber2019negative}.

\subsubsection{Theoretical results} \label{sec:asymptotic}

The output of Algorithm \ref{alg:ABC-SMC} is $(\theta_i)_{1 \leq i \leq N}$.
These can be used to produce Monte Carlo estimates of posterior quantities.
That is, given a function $h(\theta)$ then
\[
    \hat{h}_N = \frac{1}{N} \sum_{i=1}^N h(\theta_i)
\]
is a Monte Carlo estimate of the expectation of $h(\theta)$ under the ABC posterior \eqref{eq:ABC_posterior}.

The SMC sampler framework in \cite{del2006sequential} provides a proof of consistency for $\hat{h}_N$,
as well as a central limit theorem.
However the result requires the details of each SMC iteration to be prespecified.
This is not the case for ABC-SMC which adaptively selects its MCMC kernels and tolerance thresholds.
While \cite{del2012adaptive} sketch a proof for consistency properties in the adaptive setting,
\cite{beskos2016on} offer a more comprehensive asymptotic theory.
They prove the consistency of SMC estimators with adaptive tempering and Markov kernels
-- which covers the case of ABC-SMC --
providing both a weak law of large numbers and a central limit theorem,
under appropriate regularity conditions.

\subsubsection{Algorithm variations} \label{sec:ABCSMCvariations}

Unlike the ABC-SMC algorithm we have presented in Algorithm \ref{alg:ABC-SMC},
\cite{del2012adaptive} allow $r$ simulations for each particle,
$(y^t_{i,j})_{1 \leq j \leq r}$,
with weights $\tfrac{1}{r} \sum_{j=1}^r \mathbbm{1}[y^t_{i,j} \in B_{\epsilon_t}]$.
Following \citet{lee2012choice} and \citet{bernton2019approximate}, we use $r=1$,
and multiple simulations are instead performed through the design of the MCMC kernel.

As in \cite{del2012adaptive}, we generate weights $w_i^t$ using an indicator function.
A variation is to replace the indicator function with a smooth function
(and similarly replace it in \eqref{eq:ABC_jointposterior} for the associated target density).
Many of the kernel families discussed below do not apply to this setting.
Extending them to this case is an interesting possibility for future work.

\section{Kernel families} \label{sec:kernel_families}

Here we outline the different families of ABC-MCMC kernel we consider.
As discussed in Section \ref{sec:abc_intro}, the kernel acts on $(\theta, y)$ pairs
and should leave the target distribution \eqref{eq:ABC_jointposterior} invariant.
Each kernel family makes use of a proposal kernel $q(\theta' | \theta)$ whenever a parameter value is proposed.
We discuss choices of $q$ in Section \ref{sec:proposals}.

Below, Section \ref{sec:existing_families} reviews existing kernel families
and Section \ref{sec:one_hit_ind} introduces a novel kernel family
for the special case where $q$ is an independence proposal.
Section \ref{sec:early_rejection} discusses early rejection, which is implemented by some kernel families.
Section \ref{sec:mixing} reviews and discusses theoretical results on the mixing properties of these kernels.
Later sections of the paper investigate their empirical performance.

\subsection{Existing kernel families} \label{sec:existing_families}

\subsubsection{ABC Metropolis-Hastings (ABC-MH)}

Algorithm \ref{alg:ABCMH} outlines a simple kernel family.
This is essentially a Metropolis-Hastings kernel with target \eqref{eq:ABC_jointposterior}
and proposal $q(\theta' | \theta) f(y' | \theta')$.
We present the early rejection implementation of \citet{picchini2014inference}.
This changes the order of operations from the original ABC-MCMC paper \citep{marjoram2003markov},
giving a mathematically equivalent kernel but a computational speed-up.
We comment more on early rejection in Section \ref{sec:early_rejection}.
See \citet{sisson2011likelihood} for a review of ABC-MH and some extensions.

\begin{algorithm}[htbp]
    \caption{ABC Metropolis-Hastings (ABC-MH) kernel} \label{alg:ABCMH}
    \begin{algorithmic}[1]
        \Statex Input: state $(\theta, y)$; proposal kernel $q(\cdot | \theta)$
        \State Sample $\theta' \sim q(\cdot | \theta)$.
        \State Calculate $\alpha = \frac{\pi(\theta') q(\theta | \theta')}{\pi(\theta) q(\theta' | \theta)}$.
        \State With probability $1 - \min(1,\alpha)$ return $(\theta, y)$ (rejection).
        \State Sample $y' \sim f(\cdot, \theta')$
        \State If $y' \in B_\epsilon$ return $(\theta', y')$ (acceptance)
        otherwise return $(\theta, y)$ (rejection).
    \end{algorithmic}
\end{algorithm}

Algorithm \ref{alg:ABCMH} simulates a single dataset $y'$.
When $\Pr(y' \in B_\epsilon)$ is low,
this results in a high probability of leaving the state unchanged,
which is undesirable within ABC-SMC.
One solution is to repeat the kernel $r$ times,
or use a variant that performs $r$ simulations under $\theta'$ \citep{sisson2011likelihood}
(which is very similar to using multiple simulations for each particle in ABC-SMC,
as discussed in Section \ref{sec:ABCSMCvariations}).
However this requires tuning $r$.
The following kernels address this problem.

\subsubsection{One-hit} \label{sec:one_hit}

Algorithm \ref{alg:1hit} is the one-hit kernel \citep{lee2012choice}.
This can be viewed as adaptively selecting the number of simulations to perform for each $\theta'$.

\begin{algorithm}[htbp]
    \caption{One-hit kernel} \label{alg:1hit}
    \begin{algorithmic}[1]
        \Statex Input: state $(\theta, y)$; proposal kernel $q(\cdot | \theta)$
        \State Sample $\theta' \sim q(\cdot | \theta)$.
        \State Calculate $\alpha = \frac{\pi(\theta') q(\theta | \theta')}{\pi(\theta) q(\theta' | \theta)}$.
        \State With probability $1-\min(1,\alpha)$ return $(\theta, y)$ (rejection).
        \Loop
        \State Simulate $y' \sim f(\cdot | \theta')$. If $y' \in B_\epsilon$ return $(\theta', y')$ (acceptance).
        \State Simulate $y'' \sim f(\cdot | \theta)$. If $y'' \in B_\epsilon$ return $(\theta, y)$ (rejection).
        \EndLoop
    \end{algorithmic}
\end{algorithm}

A potential drawback of the one-hit kernel is that low $\alpha$ results in likely rejection
even if the likelihood at $\theta'$ is much  higher than at $\theta$.
Using counts of simulations required for $r>1$ acceptances can provide more information on this likelihood ratio.
The following kernel families implement this idea.

\subsubsection{Single proposal $r$-hit} \label{sec:rhit}

Algorithm \ref{alg:rhit} is the single proposal $r$-hit kernel of \citet{lee2012choice}.

\begin{algorithm}[htbp]
    \caption{$r$-hit kernel with single proposal} \label{alg:rhit}
    \begin{algorithmic}[1]
        \Statex Input: state $(\theta, y)$; proposal kernel $q(\cdot | \theta)$; number of proposals $r>1$
        \State Sample $\theta' \sim q(\cdot | \theta)$.
        \State Let $P = \emptyset$ and $M' = M'' = N' = N'' = 0$.
        \While{$M' < r$}
        \State Increment $N'$ by 1.
        \State Simulate $y' \sim f(\cdot | \theta')$.
        \State If $y' \in B_\epsilon$, append $y'$ to $P$ and increment $M'$ by 1.
        \EndWhile
        \State Uniformly sample $y^*$ from $P$.
        \While{$M'' < r-1$}
        \State Increment $N''$ by 1.
        \State Simulate $y'' \sim f(\cdot | \theta)$.
        \State If $y'' \in B_\epsilon$, increment $M''$ by 1.
        \EndWhile
        \State Calculate
        \[
            \alpha = \frac{\pi(\theta') q(\theta | \theta')}{\pi(\theta) q(\theta' | \theta)} \frac{N''}{N'-1}.
        \]
        \State With probability $\min(1,\alpha)$ return $(\theta',y^*)$ (acceptance).
        Otherwise return $(\theta, y)$ (rejection).
    \end{algorithmic}
\end{algorithm}

A potential drawback of this approach is if $\theta'$ has a low acceptance probability.
Then a large number of simulations could be needed until the acceptance decision is reached,
which seems computationally inefficient.
This motivates the next kernel family,
which makes multiple $\theta'$ proposals within a single acceptance decision.

\subsubsection{Multiple proposal $r$-hit}
\label{sec:multi_rhit}

Algorithm \ref{alg:multi_rhit} is the $r$-hit kernel with multiple proposals of \citet{lee2012choice}.
The implementation of ABC-SMC in \citet{bernton2019approximate} uses this kernel with $r=2$.

\begin{algorithm}[htbp]
    \caption{$r$-hit kernel with multiple proposals} \label{alg:multi_rhit}
    \begin{algorithmic}[1]
        \Statex Input: state $(\theta, y)$; proposal kernel $q(\cdot | \theta)$; number of proposals $r>1$
        \State Let $P = \emptyset$, $Q = \emptyset$ and $N' = N'' = 0$.
        \While{$|P| < r$}
        \State Increment $N'$ by 1.
        \State Sample $\theta' \sim q(\cdot | \theta)$.
        \State Simulate $y' \sim f(\cdot | \theta')$.
        \State If $y' \in B_\epsilon$, append $(\theta', y')$ to $P$.
        \EndWhile
        \State Uniformly sample $(\theta^*, y^*)$ from $P$.
        \While{$|Q| < r-1$}
        \State Increment $N''$ by 1.
        \State Sample $\theta'' \sim q(\cdot | \theta^*)$.
        \State Simulate $y'' \sim f(\cdot | \theta'')$.
        \State If $y'' \in B_\epsilon$, append $(\theta'', y'')$ to $Q$.
        \EndWhile
        \State Calculate
        \[
            \alpha = \frac{\pi(\theta^*) q(\theta | \theta^*)}{\pi(\theta) q(\theta^* | \theta)} \frac{N''}{N'-1}.
        \]
        \State With probability $\min(1,\alpha)$ return $(\theta^*,y^*)$ (acceptance).
        Otherwise return $(\theta, y)$ (rejection).
    \end{algorithmic}
\end{algorithm}

\subsection{Novel kernel family: independence one-hit} \label{sec:one_hit_ind}

Algorithm \ref{alg:one_hit_ind} is a novel alternative to the one-hit kernel.
It can be used in the special case of an independence proposal kernel i.e.~$q(\theta' | \theta) = q(\theta')$.

The motivation is as follows.
$r$-hit kernels effectively adapt the number of simulations to the acceptance probability.
However, as well as performing simulations under $\theta'$ proposals,
they also require further simulations\footnote{
    These are the simulations in step 6 of Algorithm \ref{alg:1hit},
    step 11 of Algorithm \ref{alg:rhit} and step 12 of Algorithm \ref{alg:multi_rhit}.
} under other parameter values to ensure detailed balance.
Algorithm \ref{alg:one_hit_ind} instead only performs simulations under $\theta'$,
potentially reducing the computational cost.

\begin{algorithm}[htbp]
    \caption{One-hit independence kernel for ABC-MCMC} \label{alg:one_hit_ind}
    \begin{algorithmic}[1]
        \Statex Input: state $(\theta, y)$; proposal density $q(\theta)$
        \Repeat
        \State Sample $\theta' \sim q(\cdot)$.
        \State Sample $y' \sim f(\cdot | \theta')$.
        \Until{$y' \in B_\epsilon$}
        \State Calculate $\alpha = \frac{\pi(\theta') q(\theta)}{\pi(\theta) q(\theta')}$.
        \State With probability $\min(1,\alpha)$ return $(\theta',y')$ (acceptance).
        Otherwise return $(\theta, y)$ (rejection).
    \end{algorithmic}
\end{algorithm}

Unlike $r$-hit algorithms, Algorithm \ref{alg:one_hit_ind} is a special case of Metropolis-Hastings,
and it is elementary to verify its validity.
To see this, note that steps 1--4 are a rejection method to sample from
\[
    q(\theta, y) \propto q(\theta) f(y|\theta) \mathbbm{1}(y \in B_\epsilon).
\]
Algorithm \ref{alg:one_hit_ind} is then a Metropolis-Hastings kernel with proposal $q(\theta, y)$
and ABC-MCMC target $p_\epsilon(\theta, y | y_0)$, defined in \eqref{eq:ABC_jointposterior}.
To verify this, observe that the required Metropolis-Hastings ratio is
\begin{align*}
      & \frac{p_\epsilon(\theta', y' | y_0) q(\theta, y)}{p_\epsilon(\theta, y | y_0) q(\theta', y')}                                                                                                                                           \\
    = & \frac{\pi(\theta') f(y' | \theta') \mathbbm{1}(y' \in B_\epsilon) q(\theta) f(y|\theta) \mathbbm{1}(y \in B_\epsilon)}{\pi(\theta) f(y | \theta) \mathbbm{1}(y \in B_\epsilon) q(\theta') f(y'|\theta') \mathbbm{1}(y' \in B_\epsilon)} \\
    = & \frac{\pi(\theta') q(\theta)}{\pi(\theta) q(\theta')},
\end{align*}
as used in step 5.
Therefore the kernel satisfies detailed balance
and has the correct stationary distribution \eqref{eq:ABC_jointposterior}.

(Note that $y \not \in B_\epsilon$ is transient under this kernel, and is not in the support of the target.
Therefore it suffices to consider the case $y \in B_\epsilon$.
This avoids division by zero issues in the calculation above.)

\subsection{Early rejection} \label{sec:early_rejection}

Above we noted that Algorithm \ref{alg:ABCMH} implements \textbf{early rejection}.
The idea is that steps 2 and 3 test for rejection using $\theta$ and $\theta'$ alone,
before simulating a $y'$ dataset.
Hence the expense of a simulation is sometimes avoided.

This early rejection implementation is due to \citet{picchini2014inference}.
The original ABC-MCMC paper \citep{marjoram2003markov} performs the simulation before any possibility of rejection.
This gives a mathematically equivalent kernel, but a slower implementation.

Algorithm \ref{alg:1hit} uses a similar early rejection approach,
which was contained in the original paper \citep{lee2012choice}.
However, an early rejection variation cannot easily be added to the other kernel families in Section \ref{sec:kernel_families}.
This is because the event of acceptance can no longer be represented as the intersection of two events,
one of which can be checked early because it does not involve simulations.
For instance, in Algorithm \ref{alg:one_hit_ind}
we cannot evaluate $\alpha$ early because we do not know the proposal $\theta'$
until after making the simulations.
Related delayed acceptance methods \citep{banterle2015accelerating} are sometimes possible,
but do not give a mathematically equivalent kernel.

\subsection{Mixing} \label{sec:mixing}

This section considers the mixing properties of the kernel families introduced above,
including use with an independence proposal kernel.
It is desirable to achieve geometric or uniform ergodicity if possible,
for two reasons.
Firstly, for an MCMC algorithm, the mixing properties describe the rate of convergence to the target distribution.
Intuitively it follows that poor mixing properties are also likely to make an SMC sampler produce worse approximations.
Secondly, many theoretical results on SMC assume MCMC kernels with such mixing properties
(e.g.~\citealp{whiteley2012sequential, beskos2014stability, dau2022waste}).

The theoretical mixing results covered below are summarised in Table \ref{tab:mixing}.
These properties can depend on the type of proposal kernel used.
See Section \ref{sec:proposals} for more details of particular proposal kernels.

\begin{table}[htbp]
    \begin{tabular}{c|cc}
        \multirow{2}{*}{Kernel family} & \multicolumn{2}{c}{Proposal}                             \\
                                       & Random-walk                  & Independence              \\
        \hline
                                       &                              & Yes for bounded           \\
        ABC-MH                         & No                           & $p(\theta)/q(\theta)$     \\
                                       &                              & (e.g.~defensive proposal) \\
        \midrule
        \multirow{2}{*}{One-hit}       & Yes                          & \multirow{2}{*}{Unknown}  \\
                                       & (under conditions)           &                           \\
        \midrule
        $r$-hit                        & Unknown                      & Unknown                   \\
        \midrule
        \multirow{3}{*}{\parbox{20mm}{\centering {Independence\\one-hit}}} &
        \multirow{3}{*}{\parbox{20mm}{\centering {Combination\\not possible}}} & Yes for bounded  \\
                                       &                              & $p(\theta)/q(\theta)$     \\
                                       &                              & (e.g.~defensive proposal)
    \end{tabular}
    \caption{Summary of when geometric ergodicity is known to hold.}
    \label{tab:mixing}
\end{table}

\subsubsection{ABC-MH kernel family} \label{sec:ABC_MH mixing}

Theorem 2 of \cite{lee2014variance} proves that the ABC-MH kernel family
is not geometrically ergodic under certain conditions.
One of the conditions is to have a \textbf{local proposal}.
This definition includes random walk proposals.
However, \citeauthor{lee2014variance} also note that geometric ergodicity is possible
for independence proposals, by the following argument.

Consider independence Metropolis-Hastings (IMH)
i.e.~a Metropolis-Hastings chain on $\vartheta \in \mathbb{R}^d$
with target density $p(\vartheta)$ and proposal density $q(\vartheta)$
which does not depend on the current state.
\cite{mengersen1996rates} prove that the mixing behaviour of IMH depends on
$w(\vartheta) = p(\vartheta) / q(\vartheta)$.
When $w$ is uniformly bounded above, the chain is uniformly ergodic.
Otherwise, the chain is not even geometrically ergodic.

For the ABC-MH kernel, $\vartheta = (\theta,y)$ and
\begin{align*}
    p(\vartheta) & = \pi(\theta) f(y | \theta) \mathbbm{1}[y \in B_\epsilon] / Z_p,   \\
    q(\vartheta) & = q(\theta) f(y | \theta),                                         \\
    w(\vartheta) & = \frac{\pi(\theta) \mathbbm{1}[y \in B_\epsilon]}{q(\theta) Z_p},
\end{align*}
where $Z_p>0$ is a constant and $q(\theta)$ is the independence proposal density for $\theta$.
So $w(\vartheta)$ is bounded above whenever $\pi(\theta)/q(\theta)$ is.
Section \ref{sec:defensive} below discusses one choice of proposal to ensure this.

\subsubsection{One-hit kernel family}

\cite{lee2014variance} prove geometric ergodicity of the one-hit kernel family
for local proposals under various conditions.
However the results do not extend to independence proposals,
and it remains unclear what their mixing properties are.
Also, the $r$-hit kernels are not covered by these results,
and their mixing properties are unknown.

\subsubsection{Independence one-hit kernel family}

As noted in Section \ref{sec:one_hit_ind}, this kernel can be viewed a special case of IMH with
$\vartheta = (\theta,y)$ and
\begin{align*}
    p(\vartheta) & = \pi(\theta) f(y | \theta) \mathbbm{1}[y \in B_\epsilon] / Z_p,             \\
    q(\vartheta) & = q(\theta) f(y | \theta) \mathbbm{1}[y \in B_\epsilon] / Z_q,               \\
    w(\vartheta) & = \frac{p(\vartheta)}{q(\vartheta)} = \frac{\pi(\theta) Z_q}{q(\theta) Z_p}.
\end{align*}
where $Z_p, Z_q$ are positive constants.

As noted in Section \ref{sec:ABC_MH mixing},
the mixing behaviour of IMH depends on whether there is a uniform upper bound for $w(\vartheta)$.
Later we consider training $q(\theta)$ to approximate the ABC posterior,
$\int p(\vartheta) dy$.
Taking $q(\theta)$ equal to this would give
\[
    w(\vartheta) \propto \left\{ \int f(y | \theta) \mathbbm{1}[y \in B_\epsilon] dy \right\}^{-1}
    = \Pr(y \in B_\epsilon | \theta)^{-1}
\]
which would be unbounded in many settings.
Hence the kernel would not produce geometric ergodicity.
Section \ref{sec:defensive} discusses a proposal with better mixing properties.

\section{Proposal kernels} \label{sec:proposals}

Every ABC-MCMC kernel family we consider uses a proposal kernel $q(\theta' | \theta)$ whenever a parameter value is proposed.
An ideal proposal kernel should balance exploring $\theta$ space
and making proposals likely to be accepted.

We consider some adaptive proposals trained using particles from the most recent ABC-SMC iteration.
There is a choice of exactly how to use the particles as training data.
Section \ref{sec:training_data} describes the different options.

Section \ref{sec:existing_proposal_kernels} describes existing proposal kernels:
a Gaussian random walk making local moves,
and independence proposals approximating $p_\epsilon(\theta | y_0)$
for some choice of $\epsilon$.
A motivation for the latter is to avoid wasting time simulating data for poor local proposals.
Also, independence proposals allow us to use the independence one-hit kernel
of Section \ref{sec:one_hit_ind}.
Finally, mixing guarantees vary between the different proposal kernels:
see Table \ref{tab:mixing}.

Section \ref{sec:defensive} details a defensive modification to independence proposals
which can improve their mixing properties.
Section \ref{sec:tabc} describes a novel proposal kernel using normalising flows,
and the resulting ABC-SMC algorithm, which we refer to as \textbf{transport ABC}.

\subsection{Training data} \label{sec:training_data}

In the ABC-SMC setting we can use the existing particles to train an approximation of the target distribution.
One possible training set
-- used in \cite{beaumont2009adaptive,del2012adaptive} -- is
\[
    \mathcal{A}_t = \left\{ \theta_i^{t-1} \right\}.
\]
These form an approximate sample from $p_{\epsilon_{t-1}}(\theta | y)$.
Another choice -- used by \cite{bernton2019approximate} -- is
\[
    \mathcal{B}_t = \left\{ \theta_i^{t-1} : y_i^{t-1} \in B_{\epsilon_t} \right\}.
\]
These form an approximate sample from $p_{\epsilon_t}(\theta | y)$.
This is equivalent to using the weighted $(w_i^t, \theta_i^{t-1})$ pairs,
after the particles are reweighted to $\epsilon_t$ in step 4 of Algorithm \ref{alg:ABC-SMC}.

Below we will use the notation $\mathcal{D}_t$ for a general training set,
which could be $\mathcal{A}_t$ or $\mathcal{B}_t$.
We test the choice of training set in Appendix \ref{app:target},
and find no clear advantage of either.
In the rest of the paper we use $\mathcal{B}_t$,
as it is a sample from the current target distribution,
which seems most relevant.
While $\mathcal{B}_t$ provides fewer particles to train on than $\mathcal{A}_t$,
the experiments show this does not result in significantly worse performance.

In Section \ref{sec:tabc} below we introduce training particles
$(\theta_i^{t, \text{train}}, y_i^{t, \text{train}})_{1 \leq i \leq N_\text{train}}$.
We define $\mathcal{D}_t^\text{train}$ to be one of:
\begin{align*}
\mathcal{A}_t^\text{train} &= \left\{ \theta_i^{t-1, \text{train}} \right\}, \\
\mathcal{B}_t^\text{train} &= \left\{ \theta_i^{t-1, \text{train}} : y_i^{t-1, \text{train}} \in B_{\epsilon_t} \right\}.
\end{align*}
Separate test particles are also introduced with corresponding notation.

\subsection{Existing proposal kernels} \label{sec:existing_proposal_kernels}

\subsubsection{Classic random walk} \label{sec:classicRW}

A popular proposal kernel, originally suggested by \citet{del2012adaptive}, is a Gaussian random walk
\[
    \theta' \sim \mathcal{N}(\theta, 2 \hat{\Sigma})
\]
where $\hat{\Sigma}$ is the empirical variance of $\mathcal{D}_t$.
(\citeauthor{del2012adaptive} use the training set $\mathcal{A}_t$.)
The factor of two is discussed shortly.

\subsubsection{Classic independence proposal} \label{sec:kde}

Consider the independence proposal
\begin{equation*}
    \theta'   \sim \mathcal{N}(\vartheta, 2 \hat{\Sigma}), \quad
    \vartheta \sim \mathcal{U}(\mathcal{D}_t).
\end{equation*}
This is equivalent to taking the classic random walk proposal from a uniformly sampled particle in $\mathcal{D}_t$.
This proposal is popular in ABC-PMC. 
It was introduced by \cite{beaumont2009adaptive} who motivate the covariance term (including its factor of two)
by a derivation minimising the Kullback Leibler divergence between the proposal and target distributions.
They use the training set $\mathcal{A}_t$.

\subsubsection{Mixture proposal}

\citet{bernton2019approximate} use a Gaussian mixture trained on $\mathcal{D}_t$ as an independence proposal.
(They take $\mathcal{D}_t = \mathcal{B}_t$.)
They fit the mixture using the R package MClust \citep{scrucca2023model},
which uses the EM algorithm to maximise
\begin{equation} \label{eq:cross_entropy}
    \mathcal{J}(\phi) = \sum_{\vartheta \in \mathcal{D}_t} \log q(\vartheta; \phi),
\end{equation}
where $q(\cdot; \phi)$ is the Gaussian mixture density with tuning choices $\phi$
(i.e.~weights and moments of mixture components).
So they select $\phi$ to maximise the log density on $\mathcal{D}_t$.
Also, \eqref{eq:cross_entropy} can be viewed as negative cross-entropy loss.

Our code takes a similar approach using the Python package AutoGMM \citep{athey2019autogmm}.
We generally follow the default settings.
Like MClust, these iterate through several choices
of covariance structure and other discrete tuning options,
selecting those optimising the BIC.
One choice where we depart from the defaults is that
we typically fix the number of mixture components to a prespecified value.
This is similar to \citeauthor{bernton2019approximate} who use 5 mixture components.
We investigate the number of mixture components in our experiments,
including the option of selecting this automatically.

\subsection{Defensive proposal} \label{sec:defensive}

An independence proposal density $q^*(\theta)$
can be modified to give a \textbf{defensive proposal} \citep{hesterberg1995weighted}
\begin{equation} \label{eq:defensive_q}
    q(\theta) = \eta \pi(\theta) + (1-\eta) q^*(\theta),
\end{equation}
for some small $\eta > 0$.

As discussed in Section \ref{sec:mixing}, the mixing properties of two kernel families
(ABC-MH and independence one-hit)
depend on whether the ratio $\pi(\theta) / q(\theta)$ is uniformly bounded above.
Using a defensive proposal ensures a bound of $1/\eta$.
However there is a computational trade-off due to sometimes sampling $\theta$ from the prior,
which is unlikely to be accepted for small $\epsilon$.

We investigate implementing defensive proposals in Appendix \ref{app:defensive}
and find little effect on empirical performance.
Hence improved theoretical mixing properties can be achieved if desired with little change in performance.
However, for empirical performance alone, we find no reason to use a defensive proposal.

\subsection{Novel proposal kernel: normalising flow} \label{sec:tabc}

Motivated by the mixture proposal,
in this section we describe a novel proposal kernel
based on an alternative density estimation method: normalising flows (NFs).
These are potentially more flexible than mixtures, but also more complex and slower to train.
Our later experiments investigate this trade-off.
Training NFs effectively requires a modified ABC-SMC algorithm,
Algorithm \ref{alg:TABC},
which we refer to as \textbf{transport ABC}.

\subsubsection{Normalising flows}

Normalising flows (NFs) represent a complex probability distribution
as an invertible transformation of a simple base distribution.
In terms of samples, $\theta = T(z; \phi)$ where $z$ is a sample from the base distribution,
typically $\mathcal{N}(0,\mathrm{I})$.
Here $T(\cdot;\phi)$ is a family of transformations parameterised by $\phi$.
Many flexible transformations have been proposed which allow for sampling of $\theta$ values
and calculation of the resulting density $q(\theta; \phi)$.
For general NF reviews see \cite{kobyzev2020normalizing, papamakarios2021normalizing}.
NFs have been successfully used in \textbf{simulation based inference} \citep{papamakarios2019sequential}
to estimate the posterior density $\pi(\theta | y)$ or the density of the data $f(y | \theta)$.

\subsubsection{Transport ABC}

A naive approach to generating a NF proposal would be to optimise \eqref{eq:cross_entropy}.
This can be done by stochastic gradient methods such as the Adam algorithm \citep{kingma2015adam},
which we use below.

However, naive maximisation of $\mathcal{J}(\phi)$ can \textbf{over-fit} to the training values.
A standard approach to avoid this is \textbf{early stopping}:
optimisation terminates once a similar objective function
on a separate test set stops increasing.

To apply early stopping in a SMC context, we follow \citet{arbel2021annealed},
who propose a SMC algorithm with separate train and test particle populations\footnote{
    They also use a third population -- validation particles -- to help produce an unbiased estimate of the normalising constant.
    Such an estimate is not used in the ABC context, so we omit these particles.
    (Note we follow the statistics literature naming convention for these sets.
    \cite{arbel2021annealed} follow the machine learning convention,
    which swaps the names of the test and validation sets.)}.
We train the normalising flow by maximising \eqref{eq:cross_entropy} on $\mathcal{D}^\text{train}$ (the training objective).
We also keep track of \eqref{eq:cross_entropy} on $\mathcal{D}^\text{test}$ (the test objective).
We return the flow parameters which maximise the test objective.
This avoids over-fitting.

Algorithm \ref{alg:TABC} is the resulting ``transport ABC'' algorithm.
By the theoretical results of \cite{beskos2016on} on adaptive SMC algorithms
(and assuming their regularity conditions),
discussed in Section \ref{sec:asymptotic}, the output produces consistent Monte Carlo estimates targeting the ABC posterior.
Implementation details on algorithm termination are unchanged from Section \ref{sec:termination}.
Next we discuss other implementation details.

\begin{algorithm}[htbp]
    \caption{Transport ABC} \label{alg:TABC}
    \begin{algorithmic}[1]
        \Statex Input: numbers of train and test particles $N_{\train}$, $N_{\test}$; choice of ABC-MCMC kernel
        \State Initialise $\epsilon^0 = \infty$.
        \For{$a \in \{ \text{train, test} \}$}
        \State $(\theta^{0,a}_i, y^{0,a}_i)_{1 \leq i \leq N_a} \sim \pi(\theta) f(y|\theta)$ (independent samples).
        \EndFor
        \For{$t=1,2,3,\ldots$}
        \State Calculate $\epsilon^t$ (see Section \ref{sec:threshold2}).
        \State For $a \in \{ \text{train, test} \}$, calculate weights $w^{t,a}_i = \mathbbm{1}[y^{t-1,a}_i \in B_{\epsilon_t}]$.
        \State For $a \in \{ \text{train, test} \}$, resample particles (see Section \ref{sec:resampling2}) to produce
        $(\theta'^{t-1, a}_i, y'^{t-1, a}_i)_{1 \leq i \leq N_a}$.
        \State Train $q_t(\theta)$ to estimate density of $\mathcal{D}_t^\text{train}$ using $\mathcal{D}_t^\text{test}$ for early stopping.
        \For{$a \in \{ \text{train, test} \}$}
        \For{$i=1,2,\ldots,N_a$}
        \State Apply ABC-MCMC kernel to $(\theta'^{t-1,a}_i, y'^{t-1,a}_i)$
        to output $(\theta^{t,a}_i, y^{t,a}_i)$.
        \EndFor
        \EndFor
        \EndFor
    \end{algorithmic}
\end{algorithm}

\subsubsection{Initialisation of $\phi$}

To reduce training time, we initialise $\phi$ (the weights controlling the NF shape) to a reasonable value.
For $t=1$ this is done by training $\phi$ to fit samples from the prior.
For $t>1$ we initialise $\phi$ at its final value from the previous iteration.

\subsubsection{Updating $\epsilon$} \label{sec:threshold2}

We apply our earlier methodology (see Section \ref{sec:threshold})
to each particle population to produce candidate threshold values
$\epsilon_{t, \train}$ and $\epsilon_{t, \test}$.
We then take $\epsilon_t = \max(\epsilon_{t, \train}, \epsilon_{t, \test})$.
The motivation is ensuring both populations have reasonable particle diversity.

\subsubsection{Resampling} \label{sec:resampling2}

We resample each population of particles separately.
So, for $a \in \{ \text{train, test} \}$,
we resample $(\theta'^{t-1,a}, y'^{t-1,a})_{1 \leq i \leq N_a}$ using weights $(w^{t,a}_i)_{1 \leq i \leq N_a}$
to produce $(\theta'^{t-1,a}, y'^{t-1,a})_{1 \leq i \leq N_a}$.
Systematic resampling is used, as in Section \ref{sec:resampling}.

\subsubsection{NF architecture} \label{sec:spline_flow}

Our default choice of architecture is a rational quadratic spline flow \citep{durkan2019neural}.
As tuning choices we use 50 bins, linear tails and a tail bound of 10.
The neural network generating the spline
has an initial fully connected layer with 20 hidden features, followed by 2 residual blocks.
Later we compare different types of flow (see Section \ref{sec:architecture}).

\section{Illustrations} \label{sec:illustrations}

This section illustrates some typical behaviour of the different kernels in ABC-SMC, on a stylised example.

\subsection{Quadratic model} \label{sec:quadratic}
Consider the model
\[
    y \sim \mathcal{N}(\theta_1 - \theta_2^2, 10^{-4}),
\]
with independent $\mathcal{N}(0,1)$ priors for $\theta_1$ and $\theta_2$.
As observed data we use $y_0 = 0$.
The posterior lies close to the quadratic curve $\theta_1 = \theta_2^2$, and so is highly non-Gaussian.
Therefore we expect it will be challenging
for classic proposal kernels based on a Gaussian distribution.

\subsection{Tuning choices} \label{sec:illustration_tuning}

Throughout this section we use 1000 particles in ABC-SMC.
For transport ABC this is split into 900 train particles and 100 test particles.
When training proposals we use $\mathcal{D}_t = \mathcal{B}_t$.
Another fixed tuning choice is taking $\omega = 0.5$,
the target fraction of unique particles used when updating the threshold
(see Section \ref{sec:threshold}).
Finally, in this section we use 5 components for the mixture proposal,
following \citep{bernton2019approximate}.
Appendix \ref{app:tuning} summarises our tuning choices,
including those discussed earlier in the paper.

\subsection{Approximate posterior and proposals}

\begin{figure*}[htbp]
    \includegraphics[width=0.19\mytextwidth]{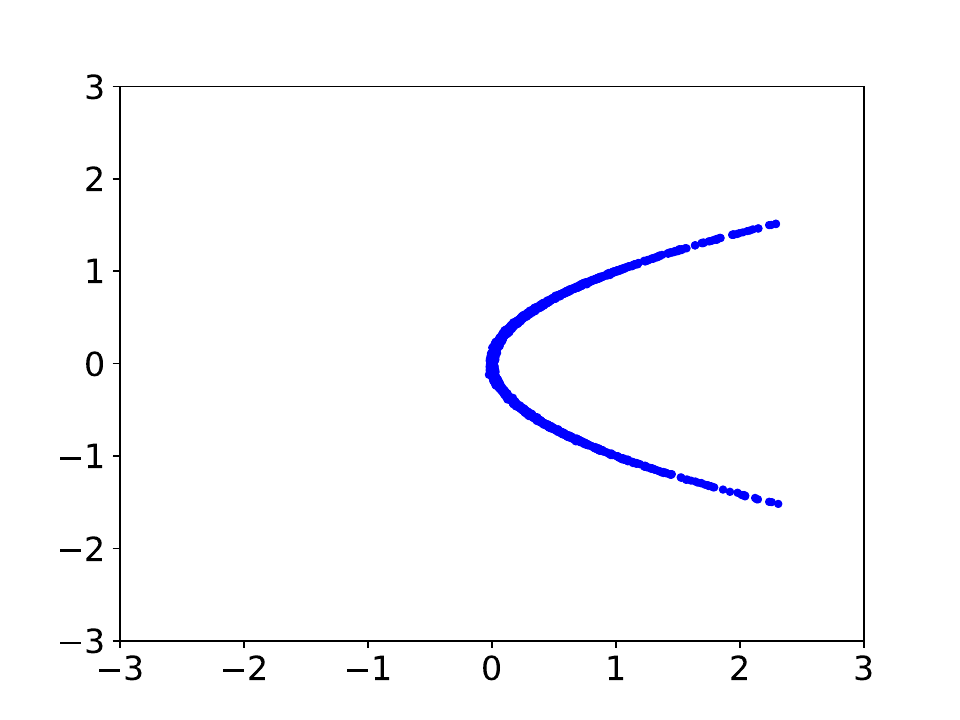}
    \includegraphics[width=0.19\mytextwidth]{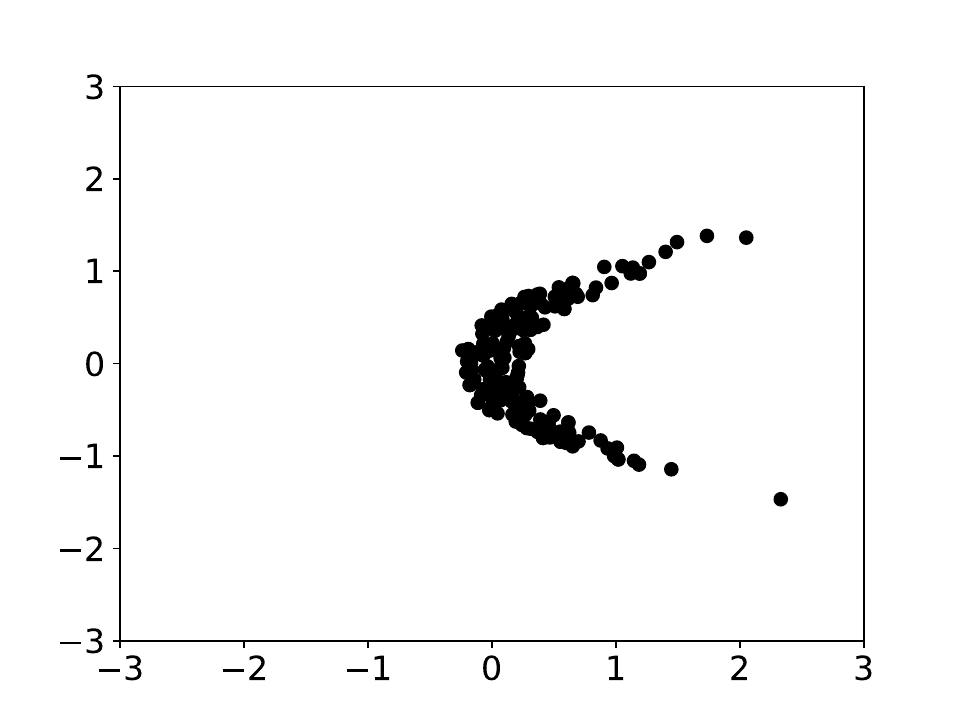}
    \includegraphics[width=0.19\mytextwidth]{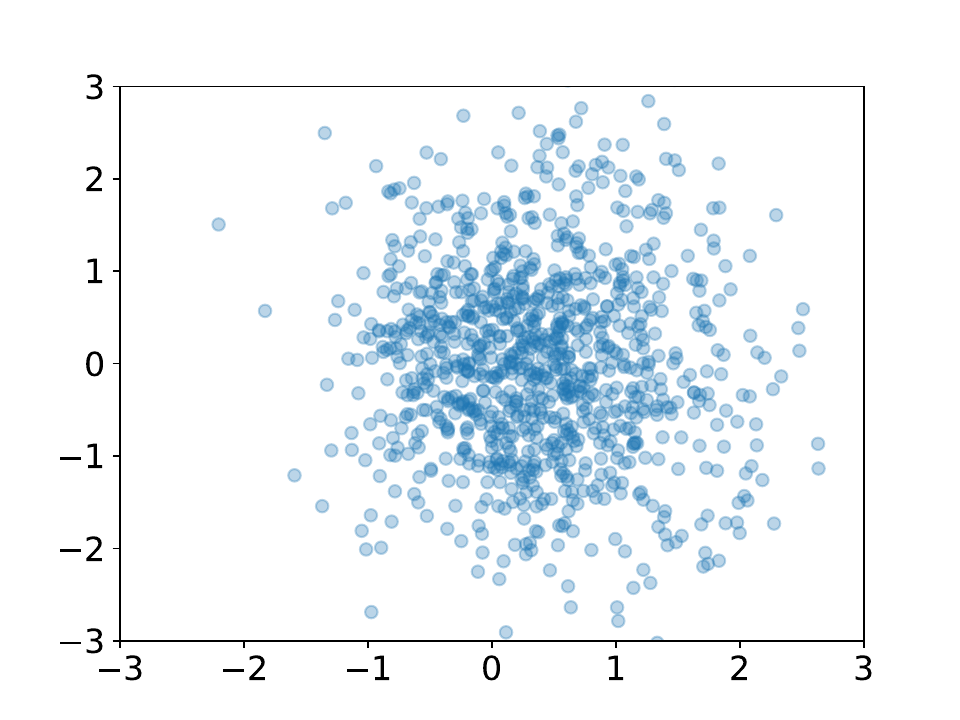}
    \includegraphics[width=0.19\mytextwidth]{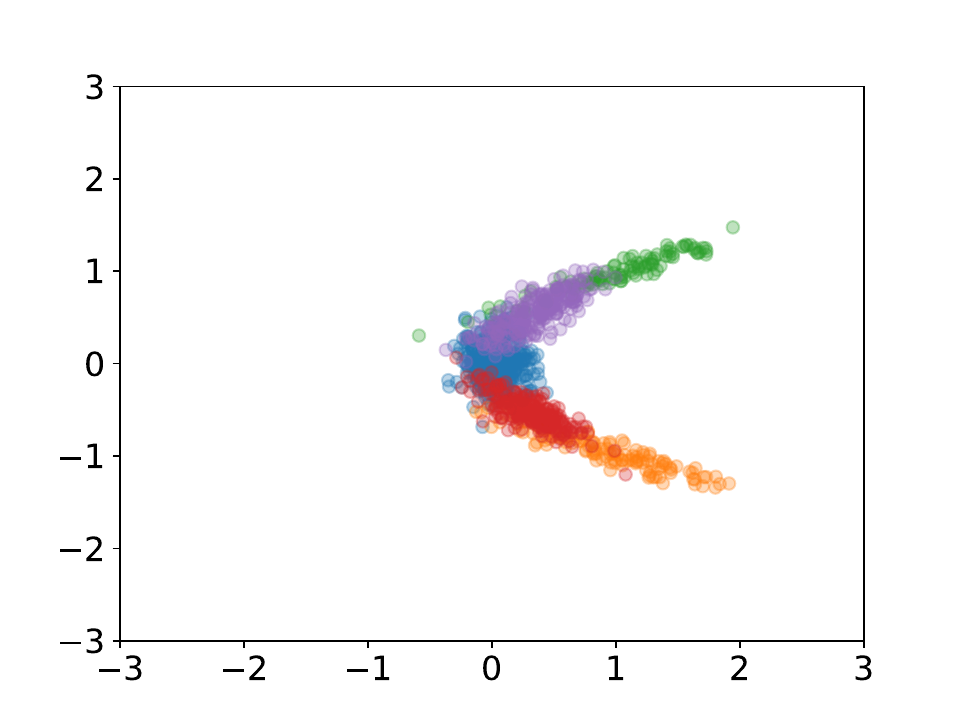}
    \includegraphics[width=0.19\mytextwidth]{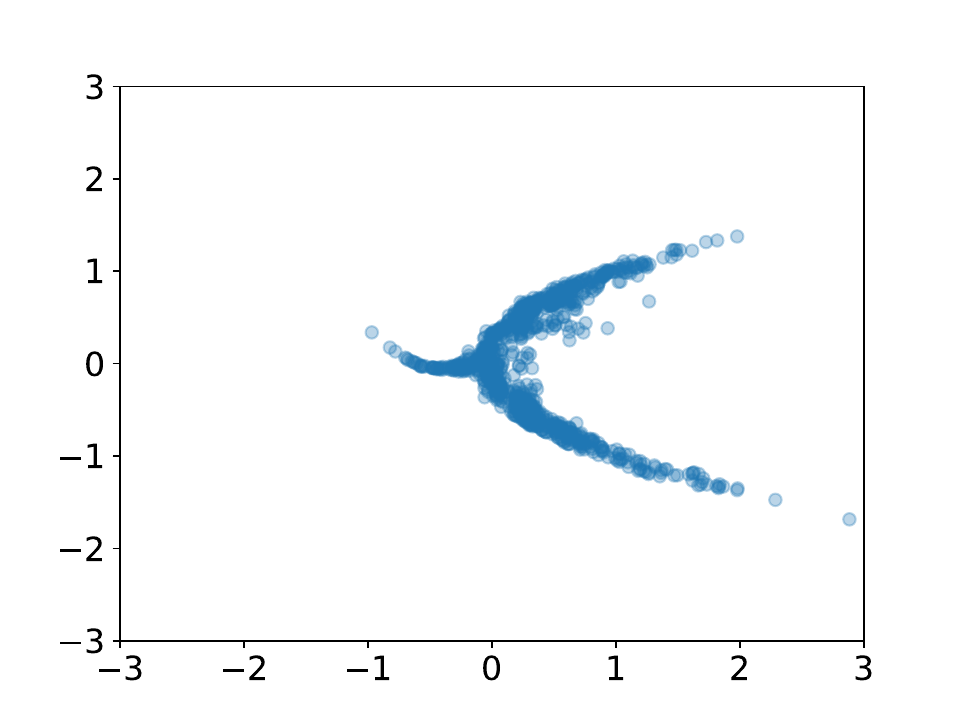}
    \caption{
        Illustrative plots of the quadratic model.
        From left to right:
        MCMC samples from the posterior,
        ABC samples from a posterior approximation,
        classic independence proposals,
        mixture proposals (different colours indicating the 5 components),
        and normalising flow proposals.
        The proposal distributions are all trained on the ABC samples.
    } \label{fig:quad_illustration}
\end{figure*}

Figure \ref{fig:quad_illustration} illustrates the ABC posterior
and various proposals for this example.
The first panel shows near-exact posterior samples based on MCMC.
These have a parabolic shape.
The second panel shows ABC samples after 3 ABC-SMC iterations.
(We use only 3 iterations so the ABC samples are clearly visually different from the MCMC samples.)
This has a wider crescent shape, reflecting the approximation error.
The remaining panels show proposals trained on the ABC samples.
The classic independence proposal covers the training data but is diffuse:
it resembles a Gaussian and hence cannot fit the crescent shape well.
The mixture proposal matches the ABC samples reasonably well,
but produces a slightly wider shape.
The flow proposal is narrower,
but has some artefacts that do not match the crescent shape:
presumably due to the relatively small training data size.

\subsection{ABC-SMC diagnostic plots}

\begin{figure*}
    \includegraphics[width=0.48\mytextwidth]{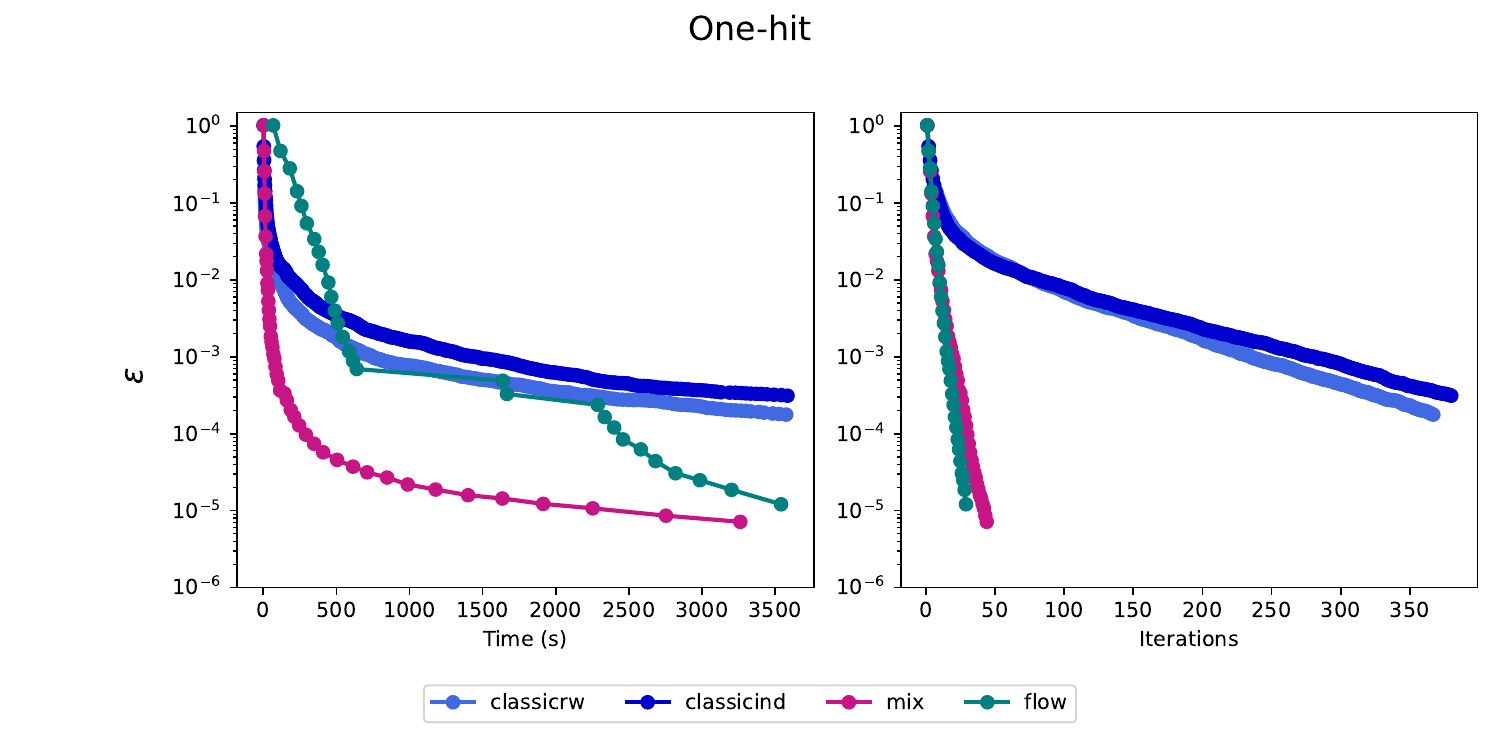}
    \includegraphics[width=0.48\mytextwidth]{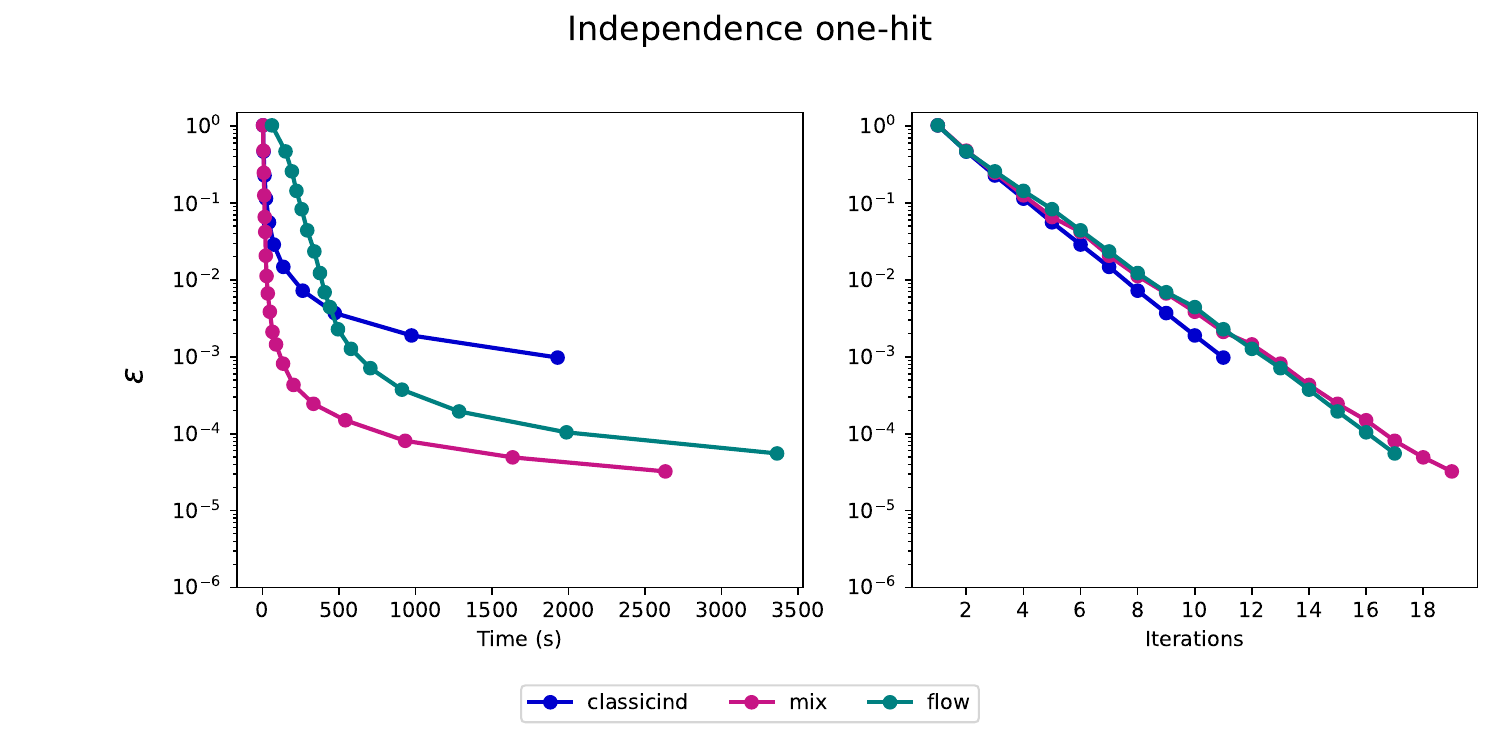} \\
    \includegraphics[width=0.48\mytextwidth]{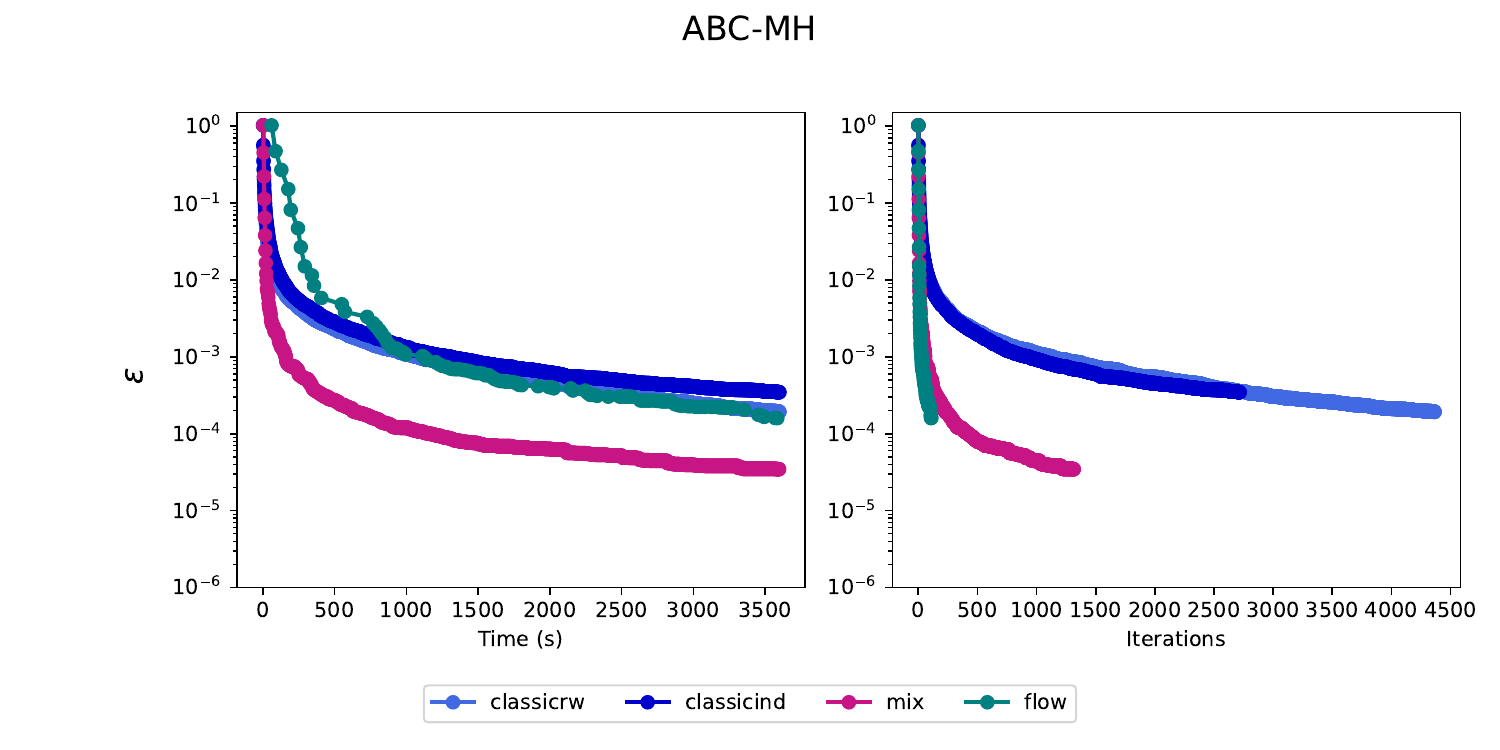}
    \includegraphics[width=0.48\mytextwidth]{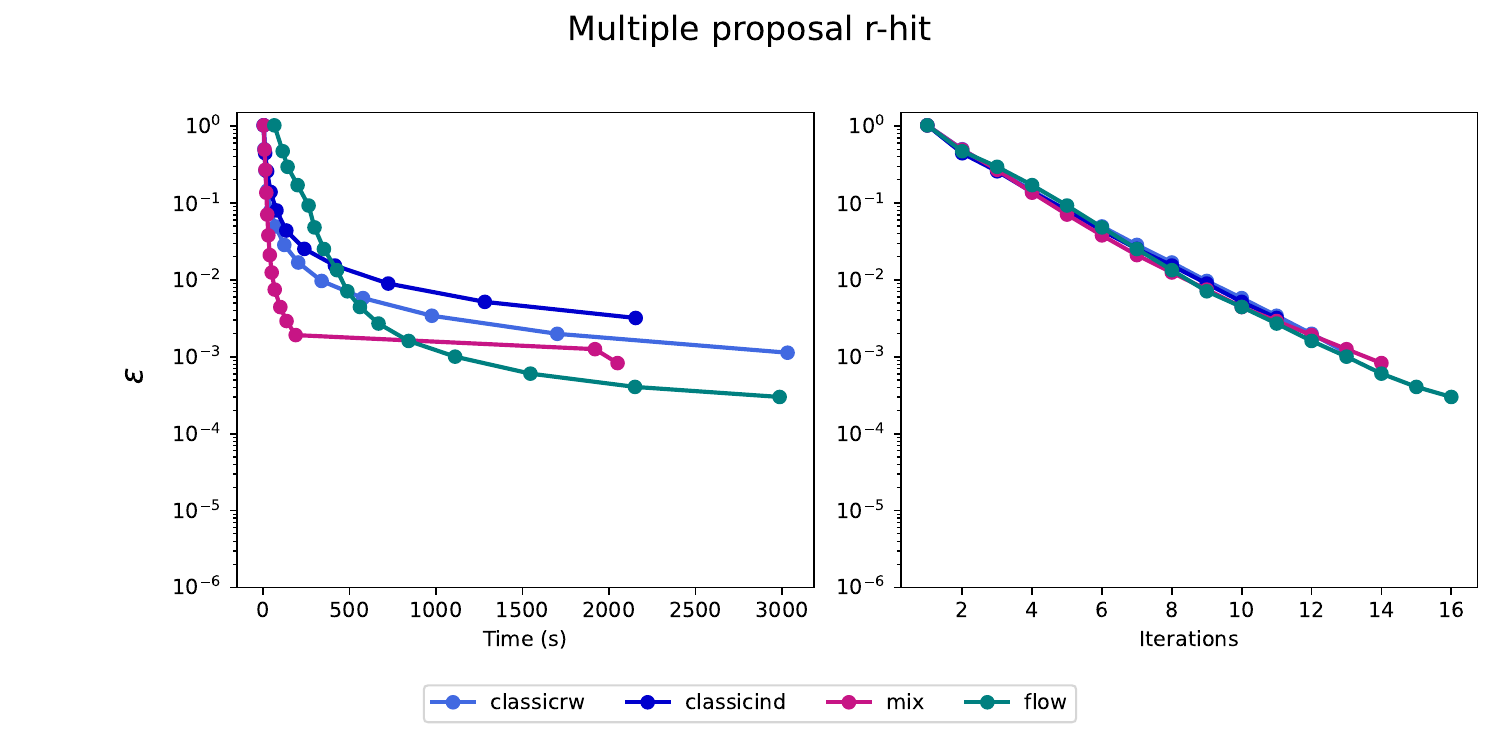}

    \caption{\label{fig:hist_eps} Quadratic model illustration of $\epsilon$ values in log scale over time (left) and iterations (right) for each kernel family.
        Note the independence one-hit kernel family requires an independence proposal,
        so the classic random walk proposal is not used.}
\end{figure*}

\begin{figure*}
    \includegraphics[width=0.24\mytextwidth]{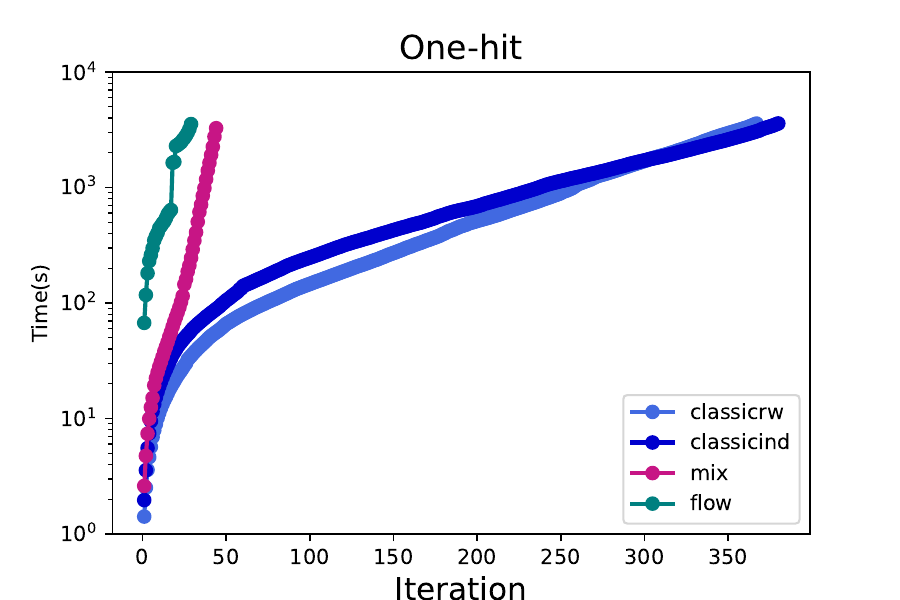}
    \includegraphics[width=0.24\mytextwidth]{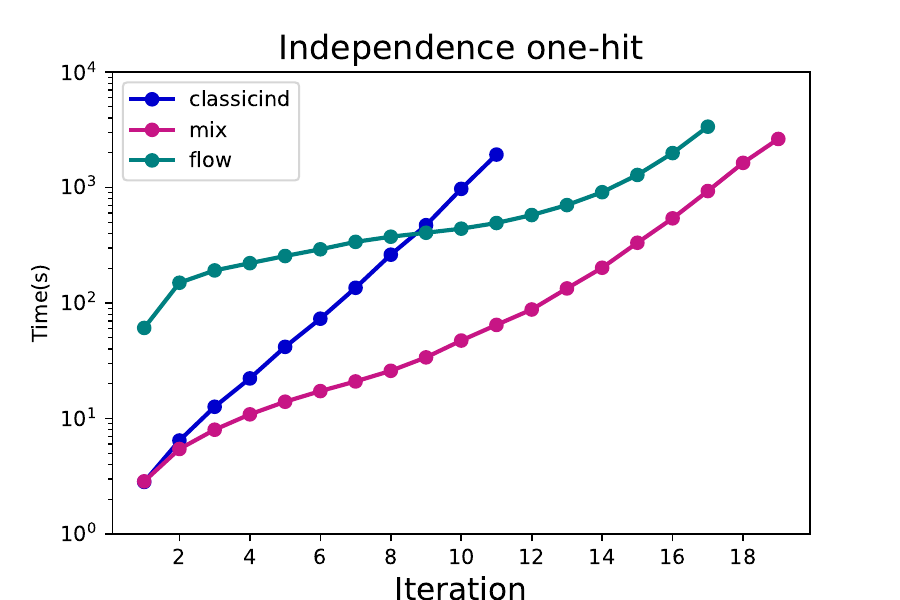}
    \includegraphics[width=0.24\mytextwidth]{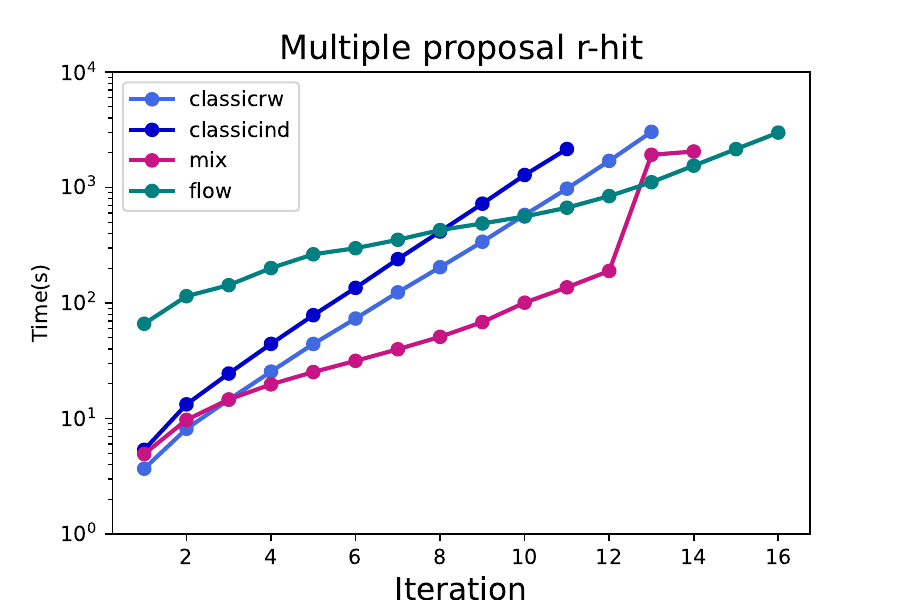}
    \includegraphics[width=0.24\mytextwidth]{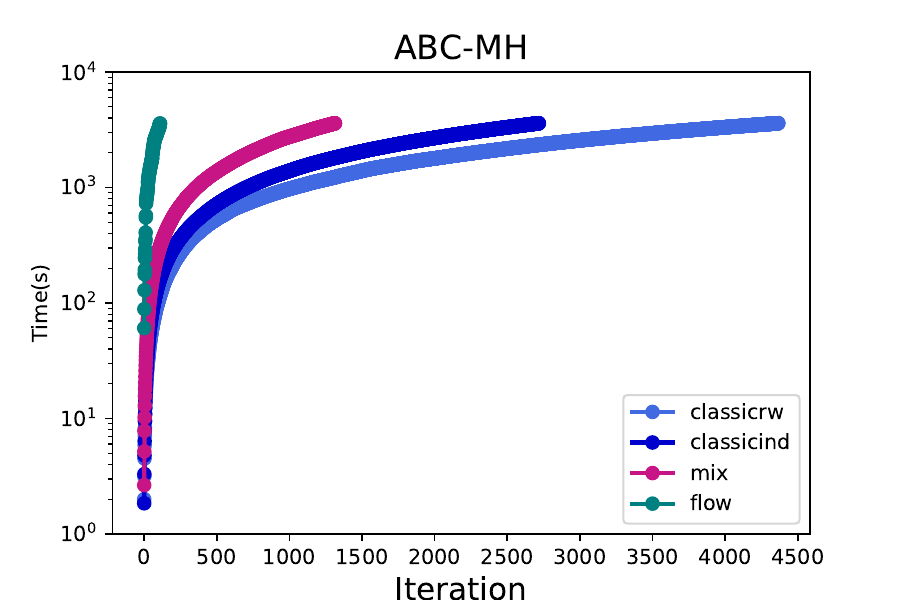}
    \caption{\label{fig:time_vs_it} Quadratic model illustration of time in log scale over iterations for each kernel family.
        Note the independence one-hit kernel family requires an independence proposal,
        so the classic random walk proposal is not used.}
\end{figure*}

\begin{figure*}
    \includegraphics[width=0.24\mytextwidth]{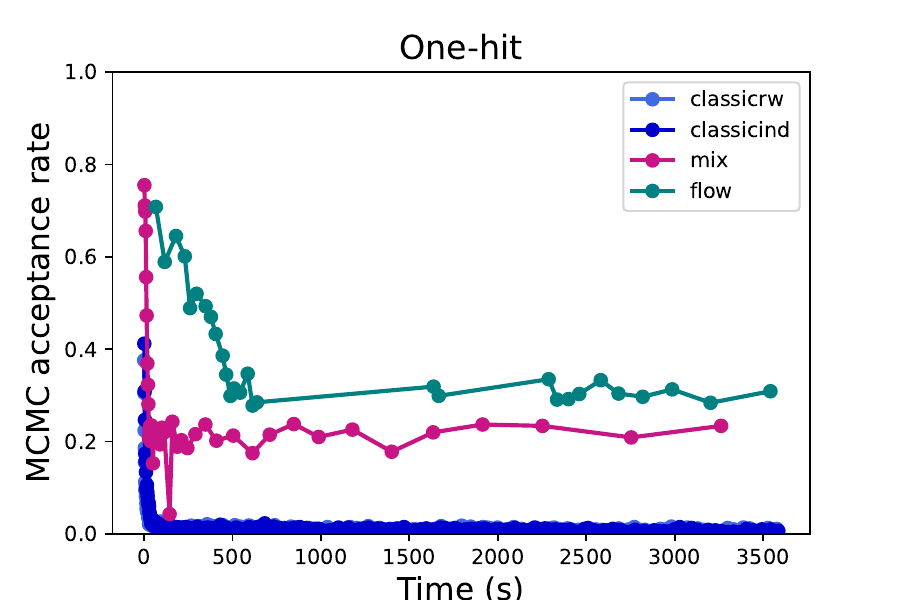}
    \includegraphics[width=0.24\mytextwidth]{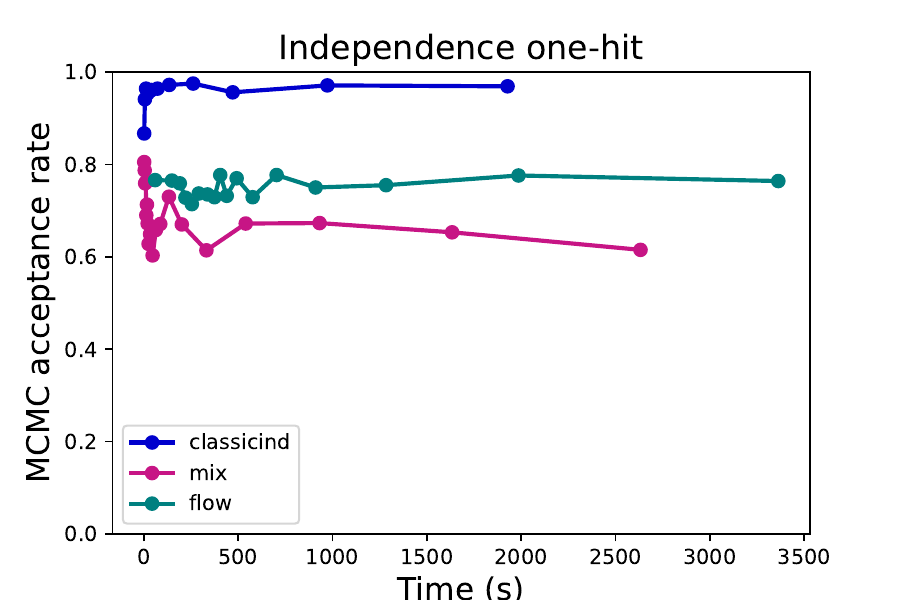}
    \includegraphics[width=0.24\mytextwidth]{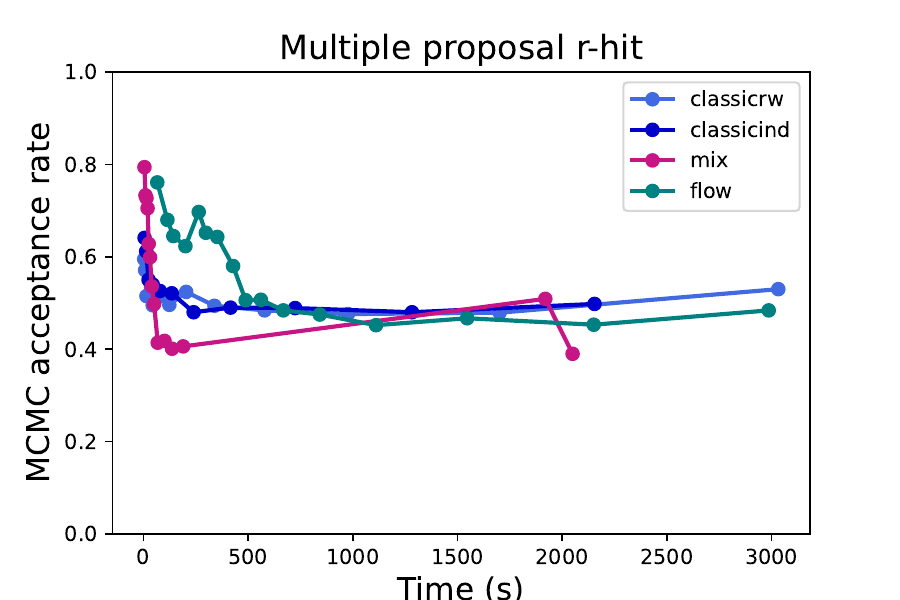}
    \includegraphics[width=0.24\mytextwidth]{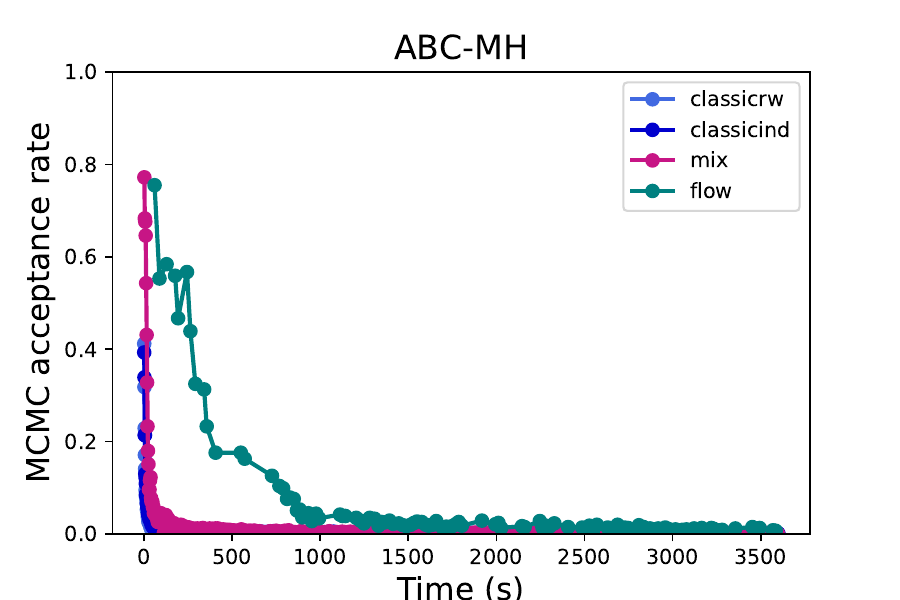}
    \caption{\label{fig:histories_time_acc} Quadratic model illustration of MCMC acceptance rate over time for each kernel families.
        Note the independence one-hit kernel family requires an independence proposal,
        so the classic random walk proposal is not used.}
\end{figure*}

Figure \ref{fig:hist_eps} shows ABC-SMC trace plots for $\epsilon$ values over time and iteration count.
This uses $\epsilon$ as a proxy for quality of the ABC-SMC output,
which is discussed in more detail in Section \ref{sec:diagnostics}.
The figure shows that the mixture proposal generally reduces $\epsilon$ fastest in terms of time.
In terms of iteration count, the flow proposal has a slight advantage over mixture in 3 out of 4 kernel families.

This illustrates that for this example the extra flexibility of the flow proposal
provides a slight advantage over the mixture proposal
when training time is neglected (by just looking at iteration count).
However, once training time is considered, the mixture proposal is more efficient.
Here we have only looked at one model without any replication of analyses,
but our detailed experiments in Section \ref{sec:main_comparison} also generally support
the mixture proposal reducing $\epsilon$ fastest in terms of time,
although it is sometimes beaten by the flow proposal.

The two classic proposals perform similarly to each other,
with a slight advantage for the random walk version in terms of time.

There are a few cases of individual iterations with unusually high durations.
This could be due to MCMC kernels occasionally taking a long time to produce an accept/reject decision.
For the flow proposal, another possibility is that training took a long time to converge.

Figure \ref{fig:time_vs_it} confirms that each iteration of the flow proposal
typically takes longer than the mixture proposal.
Due to their simplicity, the classic proposals are sometimes the fastest.
However, as we have seen, this does not lead to the fastest reduction in $\epsilon$.
Indeed, reaching smaller $\epsilon$ values increases computing time required,
as the acceptance rate become smaller.

Figure \ref{fig:histories_time_acc} shows that MCMC acceptance rates become near zero after a few iterations for ABC-MH,
as this kernel family rarely accepts proposals for small $\epsilon$ values.
Similarly, the one-hit kernel family has low acceptance rates under poor proposals.
This suggests the classic proposals are poor, as their acceptance rates become low.
Both independence one-hit and multiple proposal r-hit approaches have relatively high acceptance rates
for all proposals.

Acceptance rates are generally slightly higher for flow proposals compared to mixture.
However it is hard to interpret this result.
While it might reflect flow proposals being higher quality,
it could also be partially due to flow proposals not reaching more challenging lower $\epsilon$ values.

The independence one-hit kernel family using a classic independence proposal accepts almost all proposals.
This is because its proposal is similar to the prior (see Figure \ref{fig:quad_illustration})
giving $\alpha \approx 1$ (step 5 of Algorithm \ref{alg:one_hit_ind}).
Hence this combination performs well in terms of iteration count in Figure \ref{fig:hist_eps}.
However with this proposal each MCMC iteration is likely to require many simulations,
resulting in poor performance in terms of simulation time in Figure \ref{fig:hist_eps}.

\subsection{Inefficiency of single proposal $r$-hit kernel} \label{sec:inefficiency}

Figures \ref{fig:hist_eps}-\ref{fig:histories_time_acc} exclude one kernel family discussed earlier:
the single proposal $r$-hit kernel, Algorithm \ref{alg:rhit}.
For this proposal, we ran one experiment under each kernel family for the quadratic model with $r=2$.
We found that in all cases not a single ABC-SMC iteration was complete after an hour.
Due to the inefficiency illustrated here, we exclude this kernel family from future comparisons.

The reason for this problem is as follows.
For Algorithm \ref{alg:rhit} to terminate in either acceptance or rejection,
we must simulate $y \in B_{\epsilon_t}$ successfully $r$ times given the proposed $\theta'$.
The time required can be extremely large when $\Pr(y \in B_{\epsilon_t} | \theta')$ is low
due to proposing a poor choice of $\theta'$.
In contrast, the other kernel families have much weaker requirements to terminate
e.g.~the one-hit kernel can terminate on simulating $y \in B_{\epsilon_t}$ given $\theta$ rather than $\theta'$.

\section{Experiments} \label{sec:experiments}

This section presents experiments to investigate the performance of the different kernels in ABC-SMC.
Section \ref{sec:models} describes several statistical models which we use as example inference problems.
Section \ref{sec:diagnostics} discusses the diagnostics we use in our experiments.
Section \ref{sec:main_tuning} contains several experiments on tuning choices,
with more appearing in the appendices.
Then Section \ref{sec:main_comparison} presents our main results on a comparison of kernels.

\subsection{Models} \label{sec:models}

Here we briefly describe the models we use in our experiments.
For each, we specify observed data $y_0$,
and discuss how we generated a \textbf{reference sample} of $\theta$ values
drawn from the posterior or a good approximation.
These are used to investigate the quality of the ABC-SMC output.

\subsubsection{Quadratic model}

This is the model from Section \ref{sec:quadratic}.
Our reference sample is generated using MCMC.

\subsubsection{Gaussian mixture (GM)}
In this example \citep{sisson2007sequential},
$y$ is a sample from an equally weighted mixture of two Gaussian distributions,
$\mathcal{N}(\theta,1)$ and $\mathcal{N}(\theta,0.01)$,
with prior $\theta \sim \mathcal{U}(-10,10)$.
We use $y_0 = 0$ and draw reference samples from the exact posterior.

\subsubsection{M/G/1 queue (MG1)}
The M/G/1 queue model is a common example for likelihood-free inference
\citep{fearnhead2012constructing,papamakarios2016fast}.
There is a single queue.
Times between arrivals at the back of the queue are $Exp(\theta_1)$.
On reaching the front of the queue, service times are $\mathcal{U}(\theta_2, \theta_3)$.
Only times between departures are observed.
We use a synthetic dataset of 20 observations
generated under parameter values $\theta_1 = 0.1, \theta_2 = 4, \theta_3 = 5$.
As summary statistics, we use 5 quartiles of the inter-departure times
(following \citealp{papamakarios2016fast}).
We infer the parameters under the prior
$\theta_1 \sim \mathcal{U}(0, 1/3)$, $\theta_2 \sim \mathcal{U}(0, 10)$, $\theta_3 - \theta_2 \sim \mathcal{U}(0, 10)$ (all independent).

We use $y_0$ and reference samples from \cite{prangle2023distilling}.
The reference samples were originally generated using the MCMC method of \cite{shestopaloff2014bayesian}.
This targets the posterior conditional on the full data, rather than just the summary statistics.

\subsubsection{SEIR epidemic model}
We consider a SEIR epidemic model to describe the spreading process of an infectious disease.
It divides a closed population into four compartments:
susceptible, exposed, infectious and recovered.
We aim to infer three parameters controlling rates of transition between compartments,
based on observing a time series of reported cases:
noisy observations of new infectious cases in each time period.
Full details are in Appendix \ref{app:SEIRdetails}.

We use $y_0$ simulated under known parameter values
and generate reference samples using particle MCMC \citep{andrieu2010particle}.

\subsubsection{SLCP}
The ``simple likelihood complex posterior'' model \citep{papamakarios2019sequential}
has 4 independent observations from a bivariate Gaussian distribution
with mean $(\theta_1,\theta_2)$ and covariance matrix that is a non-linear function of $(\theta_3, \theta_4, \theta_5)$.
The parameters have independent $\mathcal{U}(-3,3)$ priors.
We simulate $y_0$ under known parameter values, and generate reference samples using MCMC. 

\subsection{Diagnostics} \label{sec:diagnostics}

Our experiments use two diagnostics of ABC-SMC performance
(1) final $\epsilon$ value and
(2) Wasserstein distance \citep{villani2009optimal} (using the Euclidean metric) between the output sample and the reference sample.
We calculate the latter using the Python Optimal Transport package \citep{flamary2021pot}.

The Wasserstein distance measures how well the ABC-SMC output approximates the target distribution.
However in some settings we find it distinguishes poorly between methods.
We describe this now, based on our findings in Section \ref{sec:main_comparison}.
Firstly, when ABC produces excellent posterior approximations, the Wasserstein distance is small for all methods
and we find its observed values are dominated by Monte Carlo sampling error.
Secondly, when ABC produces poor posterior approximations, the Wasserstein distance is large for all methods.
This can occur if the summary statistics are not informative enough,
or if the problem is too challenging for ABC to produce a good posterior approximation under our time constraints.

Due to these issues we also use the final $\epsilon$ as a diagnostic.
We find this can often distinguish between methods even when the Wasserstein distance cannot.
\citet{lueckmann2021benchmarking} review other diagnostics such as classifier score,
but we expect these to behave similarly to Wasserstein distance in this setting.

\subsection{Tuning choices} \label{sec:main_tuning}

We use the same default tuning choices as for our earlier illustrations.
See Section \ref{sec:illustration_tuning} for an overview, and Appendix \ref{app:tuning} for a full list.
For experiments in this section we terminate each ABC-SMC algorithm after 1 hour,
and report results for the last complete iteration.

The remainder of this section investigates varying several tuning choices for certain ABC-SMC kernels.
This justifies many of our default choices.
Some other tuning experiments appear in the appendices.
Appendix \ref{app:target} considers the choice of training target (see Section \ref{sec:training_data}),
and Appendix \ref{app:defensive} considers defensive proposals (see Section \ref{sec:defensive}).
Appendix \ref{app:fallback} investigates whether to include a mechanism to avoid underfitting proposals.

\subsubsection{Mixture components} \label{sec:mixture_components}

Here we vary the number of components used for the mixture proposal.
Prior work \citep{bernton2019approximate} used 5 components.
We investigate 3--10 components, as well as making an automatic choice in this range
(using the default selection method in the AutoGMM package).
We performed experiments for every model, all using the one-hit kernel.
For each model we ran 5 replications, using shared random seeds
i.e.~under every model the $i$th replication uses seed $s_i$.

Automatic selection sometimes crashed.
This was the case for all SEIR replications and 2 GM replications.
So we concentrate on investigating a fixed number of components.
Nonetheless, we include results for automatic selection in our results tables.

Tables \ref{tab:components_epsilon} and \ref{tab:components_loss} shows the results.
We find that the best number of components is dependent on the model
and the diagnostic used.
Looking at $\epsilon$ values, for most models the best choice is in the range 8--10.
However for GM the best choice is 3 components.
Further plots (see Appendix \ref{app:extra_results}) support these results,
by showing clear relationships between $\epsilon$ and number of components.

Looking at loss values, the results are less clear.
For SLCP model lower choices seem better: 3 components does best.
For MG1 and SEIR, the best choice is at the higher end of the range: 7-9 components.
For GM and quadratic model, the number of components seems to have less effect,
with most of the range producing similar loss values.
Plots (see Appendix \ref{app:extra_results}) suggest the relationships between loss and number of components is weak
in most cases.

Looking at the results for an automatic choice,
it sometimes beats the best fixed number of components
for $\epsilon$ (in 2/5 models),
but does not for loss.
This relatively poor performance may be due in part to its added computational expense.
There may be scope for more efficient and stable ways to automatically make this tuning choice within an ABC-SMC algorithm:
see Section \ref{sec:conclusion} for a discussion.

Overall we conclude that the best choice of number of components is situation dependent.
So we do not offer a recommendation of how many to use in general.
The weak relationship between loss and number of components
suggests that the results are relatively robust to this tuning choice.

For our subsequent experiments, we use 5 components, matching previous work \citep{bernton2019approximate}.
This is a reasonably performing choice from our results for our particular models,
but is rarely the best.
So making this choice investigates how well the mixture kernel performs
without taking advantage of an expensive exercise in turning the number of components.

\begin{table*}
    \begin{tabular}{l|rrrrrrrr|r}
        \toprule
                  & \multicolumn{9}{c}{Mixture components}                                                                                                                          \\
        model     & 3                                      & 4               & 5       & 6               & 7       & 8               & 9               & 10             & Automatic \\
        \midrule
        GM        & \best{1.45e-4}                         & \vgood{1.72e-4} & 2.25e-4 & \vgood{1.94e-4} & 2.03e-4 & 2.64e-4         & 2.62e-4         & 2.23e-4        & 1.41e-4   \\
        MG1       & 1.39                                   & 1.35            & 1.33    & 1.31            & 1.31    & \vgood{1.31}    & \best{1.25}     & \vgood{1.28}   & 1.31      \\
        quadratic & 1.60e-5                                & 9.39e-6         & 4.55e-6 & 3.93e-6         & 3.42e-6 & \vgood{3.23e-6} & \vgood{3.08e-6} & \best{2.80e-6} & 2.37e-6   \\
        SEIR      & 25.5                                   & 25.1            & 24.7    & 24.6            & 24.6    & \vgood{24.5}    & \best{24.3}     & \vgood{24.4}   & -         \\
        SLCP      & 0.936                                  & 0.857           & 0.773   & 0.746           & 0.737   & \vgood{0.703}   & \vgood{0.681}   & \best{0.660}   & 0.716     \\
        \bottomrule
    \end{tabular}
    \caption{
        Results from varying number of mixture components.
        Final $\epsilon$ values are shown to 3 significant figures.
        In each row, the best results -- excluding automatic -- are shown as \best{0.01}, and the second and third best as \vgood{0.02}.
        Values are means across five replications, except for automatic where some replications crashed.
        The dash represents it crashing in all replications for SEIR.
        It also crashed in 2 out of 5 GM replications.
    }
    \label{tab:components_epsilon}

    \begin{tabular}{l|rrrrrrrr|r}
        \toprule
                  & \multicolumn{9}{c}{Mixture components}                                                                                                                          \\
        model     & 3                                      & 4             & 5             & 6            & 7            & 8             & 9             & 10           & Automatic \\
        \midrule
        GM        & 0.267                                  & \vgood{0.230} & \vgood{0.230} & 0.284        & 0.292        & 0.236         & 0.341         & \best{0.228} & 0.266     \\
        MG1       & 1.81                                   & 1.70          & 1.52          & \vgood{1.45} & 1.47         & 1.46          & \best{1.37}   & \vgood{1.43} & 1.47      \\
        quadratic & 0.158                                  & 0.120         & 0.142         & \best{0.0988}& \vgood{0.105}& 0.109         & \vgood{0.100} & 0.124        & 0.127     \\
        SEIR      & 0.671                                  & 0.628         & 0.551         & 0.541        & \best{0.502} & \vgood{0.507} & \vgood{0.536} & 0.600        & -         \\
        SLCP      & \best{0.818}                           & \vgood{0.914} & \vgood{0.929} & 0.996        & 0.941        & 0.985         & 1.03          & 1.02         & 0.951     \\
        \bottomrule
    \end{tabular}
    \caption{
        As Table \ref{tab:components_epsilon} but showing final Wasserstein loss values.
    }
    \label{tab:components_loss}
\end{table*}

\subsubsection{Flow architecture} \label{sec:architecture}

Normalising flows involve many tuning choices,
including the type and number of flow layers used.
In this section we test the robustness of our results to these choices.

We consider three flow architectures:
\texttt{spline} (a single rational quadratic spline layer),
\texttt{spline\_two\_layers} (adding a second layer of the same form)
and \texttt{coupling} (using 4 coupling layers instead).
Elsewhere in the paper we typically use \texttt{spline}.
For general background on these flows see \cite{kobyzev2020normalizing,papamakarios2021normalizing}.

More details on the spline layers can be found in Section \ref{sec:spline_flow}.
For \texttt{coupling}, each coupling layer's transformation is a residual network with 20 hidden features and 10 blocks.

\begin{figure*}
    \includegraphics[width=\mytextwidth]{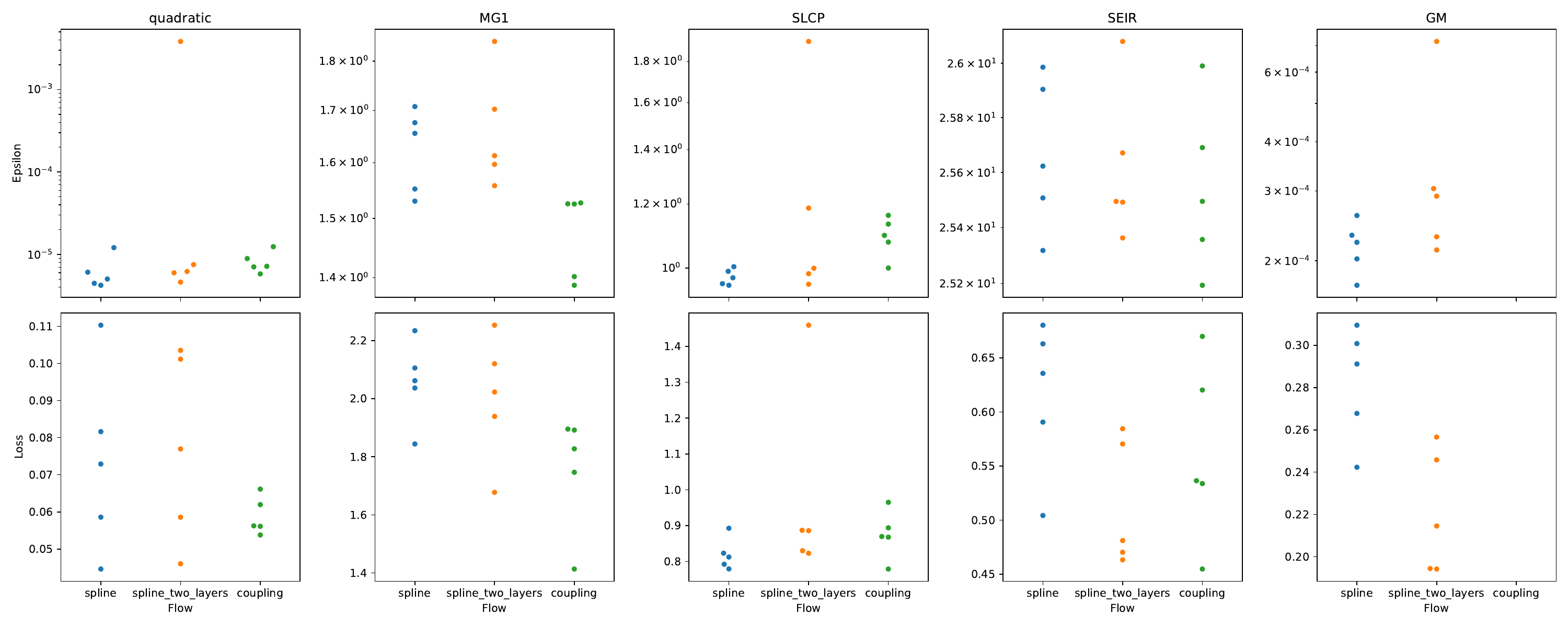}
    \caption{
        Results from varying the choice of flow:
        final $\epsilon$ values (top) and Wasserstein loss values (bottom).
        Each point represents a single experiment.
        \texttt{coupling} could not be used for the GM model -- see main text for details.
    } \label{fig:flow_results}
\end{figure*}

For each model and choice of flow we ran 5 replications using shared random seeds, as earlier.
Each used the one-hit kernel.
The results indicate the flows give similar performance.
We found results tables did not illustrate this clearly,
so instead we present the results in Figure \ref{fig:flow_results}.

Visual inspection suggests some architectures may have mild advantages for particular models.
For instance \texttt{spline\_two\_layers} has some large outlying $\epsilon$ values
for the quadratic and SLCP models.
However there is no obvious best architecture across models.
Note that a drawback of \texttt{coupling} is that it cannot be used for the Gaussian mixture model.
This is because the posterior has only one parameter and a coupling flow requires at least two.

Overall our results seem robust to the choice of flow.
In the remainder of our experiments we use the \texttt{spline} flow,
as it is simpler than \texttt{spline\_two\_layers}
and can be used for a single parameter, unlike \texttt{coupling}.

\subsubsection{Multiple proposal $r$-hit kernel}

Here we vary the choice of $r$ in the multiple proposal $r$-hit kernel, Algorithm \ref{alg:multi_rhit}.
Throughout we use a mixture proposal with 5 components.

Table \ref{tab:hits} shows the results (final $\epsilon$ and Wasserstein loss values)
for each model, taking $r$ between 2 and 5.
These are means for 5 replications.
The replications use shared random seeds, as earlier.
Increasing $r$ results in higher $\epsilon$ values.
This is due to the increased computational cost of using the kernel,
as more acceptances are required.
For three models (MG1, SEIR, SLCP), $r=2$ gives the best loss values.
For the remaining models (GM, quadratic), $r=3$ or $4$ is the best choice.
However, performing ANOVA tests shows that the choice of $r$ is not significant for these two last models
($p$-values of $0.67$, $0.74$).
It appears the differences in loss values here are driven by Monte Carlo variation.
Therefore we conclude that $r=2$ is a good tuning choice.

\begin{table*}
    \begin{tabular}{l|rrrr|rrrr}
        \toprule
                                               & \multicolumn{4}{c|}{Final $\epsilon$} & \multicolumn{4}{c}{Loss}                                                                                               \\
        \diagbox[height=1.5\line]{Model}{Hits} & 2                                     & 3                        & 4               & 5       & 2              & 3             & 4             & 5              \\
        \midrule
        GM                                     & \best{3.31e-3}                        & \vgood{5.48e-3}          & \vgood{7.39e-3} & 1.14e-2 & \vgood{0.284}  & \best{0.260}  & \vgood{0.262} & 0.307          \\
        MG1                                    & \best{2.02}                           & \vgood{2.21}             & \vgood{2.46}    & 2.60    & \best{2.26}    & \vgood{2.45}  & \vgood{2.63}  & 2.67           \\
        quadratic                              & \best{1.09e-4}                        & \vgood{1.59e-4}          & \vgood{2.12e-4} & 3.34e-4 & \vgood{0.0883} & 0.0980        & \best{0.0847} & \vgood{0.0949} \\
        SEIR                                   & \best{27.9}                           & \vgood{29.1}             & \vgood{30.9}    & 32.1    & \best{0.648}   & \vgood{0.687} & \vgood{0.693} & 0.731          \\
        SLCP                                   & \best{1.03}                           & \vgood{1.13}             & \vgood{1.16}    & 1.21    & \best{0.899}   & \vgood{0.912} & \vgood{0.944} & 0.969          \\
        \bottomrule
    \end{tabular}
    \caption{
        Results from varying $r$ in the multiple proposal $r$-hit kernel.
        The final $\epsilon$ (left) and Wasserstein loss (right) values are shown to 3 significant figures.
        In each case, the best results in a row are shown as \best{0.01}, and the second and third best as \vgood{0.02}.
    }
    \label{tab:hits}
\end{table*}

\subsection{Main comparison results} \label{sec:main_comparison}

Our main comparison compares all combinations of kernel family
(except the single proposal $r$-hit which we find can be very inefficient
-- see Section \ref{sec:inefficiency})
and proposal kernel, using the tuning choices identified above.
Five replications are run for each model, using shared random seeds as earlier.
Table \ref{tab:eps_table} shows the mean $\epsilon$ results,
and Table \ref{tab:loss_table} shows the mean Wasserstein loss results.
We also ran statistical tests on the results, described below.

\subsubsection{$\epsilon$ results}

For each model we used ANOVA to test the hypothesis that
the mean final $\epsilon$ values were the same for all combinations of kernel family and proposal considered.
In every case the hypothesis was decisively rejected ($p \leq 10^{-20}$).
This confirms that the choice of kernel has a significant effect on $\epsilon$.

The $\epsilon$ results show support for a mixture proposal.
For every model, this produces some or all of the combinations in the top 3.
The flow proposal (with the one-hit kernel) gives the best combination for two models,
but it is less consistent than the mixture proposal.

Under the mixture proposal, the ABC-MH kernel and one-hit kernel both perform best 2 times out of 5,
with the independence one-hit doing best for the remaining model.
The one-hit kernel is also usually the best choice for the flow proposal (top 4 times out of 5).
However for both classic proposals, the ABC-MH kernel is usually top (at least 4 times out of 5).
The multi-hit kernel is usually the worst choice for all proposals (worst 3-5 times out of 5 for each proposal).

\subsubsection{Loss results} \label{sec:loss_results}

For each model we used ANOVA to test the hypothesis that
the mean Wasserstein loss values were the same for all combinations of kernel family and proposal considered.
For four models the hypothesis was decisively rejected ($p \leq 10^{-3}$), but the results were less decisive for the fifth model, GM ($p=0.016$).

Looking at MG1, SEIR and SLCP, the order of the loss results are in rough agreement with the $\epsilon$ results.
These are the models where very small $\epsilon$ values are not reached.
The correspondence between $\epsilon$ and loss results suggests that here the Wasserstein loss of the output sample
reflects the quality of the ABC posterior, which is driven by $\epsilon$.

In the remaining models (GM, quadratic) the order of loss results
appears different to that of the $\epsilon$ results.
For these models, all experiments reach $\epsilon$ values below $0.01$.
So we expect that all the ABC posteriors are good approximations of the true posteriors\footnote{
    To see why, note the ABC posterior \eqref{eq:ABC_posterior}
    is the posterior for a model adding extra noise which is uniform on $B_\epsilon$ \citep{wilkinson2013approximate}.
    For both models the final $\epsilon$ values are small compared to the Gaussian scales in the data generating model.
    Thus it is intuitively reasonable that the extra noise produces little effect on the posterior.
}.
Hence it seems plausible that the ordering of these loss results is no longer driven by the value of $\epsilon$ achieved,
and other aspects of the kernel become more important.
However there is little evidence to decide on what these aspects are,
as the best combinations for GM and quadratic are quite different.
Furthermore, we argue this regime is rare in real applications:
usually it is not feasible for ABC to reduce $\epsilon$
far enough that its effect becomes negligible.

\begin{table*}[tbph]
    \scriptsize
    \begin{center}
        \begin{tabular}{l|rrrr|rrrr}
            \toprule
                      & \multicolumn{4}{c|}{classic independence} & \multicolumn{3}{c}{classic RW}                                                                                                         \\
            model     & one-hit                                   & ind\_1\_hit                    & multi\_r\_hit & ABC-MH       & one-hit          & multi\_r\_hit & ABC-MH                              \\
            \midrule
            GM        & 3.96e-04                                  & 7.34e-03                       & 2.51e-03      & 4.30e-04     & 4.26e-04         & 1.46e-03      & \vgood{2.82e-04}                    \\
            MG1       & 2.00                                      & 1.89                           & 2.64          & 1.53         & 2.67             & 2.56          & \vgood{1.34}                        \\
            quadratic & 4.28e-04                                  & 7.68e-04                       & 2.25e-03      & 3.56e-04     & 2.43e-04         & 1.38e-03      & 2.05e-04                            \\
            SEIR      & 32.1                                      & 29.0                           & 35.8          & 26.5         & 30.6             & 36.2          & 25.4                                \\
            SLCP      & 1.53                                      & 1.23                           & 1.51          & 1.18         & 1.82             & 1.56          & 1.11                                \\
            \midrule
                      & \multicolumn{4}{c|}{flow}                 & \multicolumn{4}{c}{mix}                                                                                                                \\
            model     & one-hit                                   & ind\_1\_hit                    & multi\_r\_hit & ABC-MH       & one-hit          & ind\_1\_hit   & multi\_r\_hit    & ABC-MH           \\
            \midrule
            GM        & \best{2.18e-04}                           & 2.42e-03                       & 9.05e-03      & 7.16e-03     & \vgood{2.27e-04} & 1.03e-03      & 2.99e-03         & 1.00e-03         \\
            MG1       & 1.65                                      & 2.36                           & 3.30          & 2.46         & 1.35             & \vgood{1.34}  & 1.99             & \best{1.30}      \\
            quadratic & \best{5.68e-06}                           & 7.49e-05                       & 2.66e-04      & 1.30e-04     & \vgood{5.85e-06} & 2.85e-05      & 1.09e-04         & \vgood{2.56e-05} \\
            SEIR      & 25.7                                      & 25.7                           & 29.4          & \vgood{24.8} & 24.9             & \vgood{24.6}  & 27.9             & \best{22.7}      \\
            SLCP      & 0.964                                     & 1.35                           & 1.63          & 1.36         & \vgood{0.816}    & \best{0.806}  & 1.04             & \vgood{0.828}    \\
            \bottomrule
        \end{tabular}

        \caption{Main comparison results: final $\epsilon$ values to 3 significant figures.
            For each model the best results are shown as \best{0.01}, and the second and third best as \vgood{0.02}.
            Note that each model is split across two rows.
        }
        \label{tab:eps_table}
    \end{center}

    \begin{center}
        \begin{tabular}{l|rrrr|rrrr}
            \toprule
                      & \multicolumn{4}{c|}{classic independence} & \multicolumn{3}{c}{classic RW}                                                                                                \\
            model     & one-hit                                   & ind\_1\_hit                    & multi\_r\_hit & ABC-MH       & one-hit       & multi\_r\_hit & ABC-MH                        \\
            \midrule
            GM        & 0.356                                     & 0.270                          & 0.285         & 0.247        & 0.298         & 0.273         & 0.274                         \\
            MG1       & 2.38                                      & 2.27                           & 2.81          & 1.93         & 2.80          & 2.73          & 1.55                          \\
            quadratic & 0.0772                                    & \best{0.0428}                  & \vgood{0.0487}& 0.0947       & 0.136         & \vgood{0.0541}& 0.127                         \\
            SEIR      & 0.738                                     & 0.703                          & 0.929         & 0.651        & 0.726         & 0.880         & 0.759                         \\
            SLCP      & 1.19                                      & 0.968                          & 1.14          & 1.02         & 1.37          & 1.20          & 0.992                         \\
            \midrule
                      & \multicolumn{4}{c|}{flow}                 & \multicolumn{4}{c}{mix}                                                                                                       \\
            model     & one-hit                                   & ind\_1\_hit                    & multi\_r\_hit & ABC-MH       & one-hit       & ind\_1\_hit   & multi\_r\_hit & ABC-MH        \\
            \midrule
            GM        & 0.282                                     & 0.264                          & \vgood{0.244} & \best{0.207} & \vgood{0.224} & 0.270         & 0.260         & 0.253         \\
            MG1       & 2.08                                      & 2.49                           & 2.93          & 2.53         & \vgood{1.54}  & \vgood{1.55}  & 2.22          & \best{1.27}   \\
            quadratic & 0.0724                                    & 0.0647                         & 0.0555        & 0.0897       & 0.139         & 0.103         & 0.0883        & 0.133         \\
            SEIR      & 0.615                                     & 0.574                          & 0.752         & 0.624        & \best{0.547}  & \vgood{0.559} & 0.648         & \vgood{0.553} \\
            SLCP      & \best{0.813}                              & 1.03                           & 1.23          & 1.07         & \vgood{0.916} & 0.956         & \vgood{0.914} & 0.964          \\
            \bottomrule
        \end{tabular}

        \caption{As Table \ref{tab:eps_table} but showing final Wasserstein loss values.}
        \label{tab:loss_table}
    \end{center}
\end{table*}

\subsubsection{Summary} \label{sec:summary}

Our results support the mixture proposal as a robust choice
across different kernel families and diagnostics.
Similarly the one-hit kernel is also robust choice.
Some particular combinations also perform well, such as
(1) flow proposal and one-hit kernel
(2) mixture proposal and independence one-hit kernel
(3) mixture proposal and ABC-MH kernel

However our overall recommendation is to use the one-hit kernel with a mixture proposal,
since this combination is a top performer,
and its kernel family and proposal also individually always perform reasonably well.

\section{Conclusions} \label{sec:conclusion}

Our overall recommendation is to use the one-hit kernel with a mixture proposal (see Section \ref{sec:summary}).
We find 5 mixture components behaves robustly in our experiments and gives good results,
although the best number of components is problem specific (see Section \ref{sec:mixture_components}).
\cite{bernton2019approximate} make the same choices,
except they use the multiple proposal $r$-hit kernel with $r=2$.

Our novel methods -- normalising flow proposal and independence one-hit kernel --
were often competitive but were not the best performers.
Below we discuss possible reasons for this and other aspects of our findings.

\subsection{Choice of proposal}

Our results show that the classic Gaussian proposals with variance $2 \hat{\Sigma}$
are outperformed by training proposals from more flexible families of distributions.
This is unsurprising for highly non-Gaussian target distributions,
as illustrated in Figure \ref{fig:quad_illustration}.
Interestingly we find that mixture proposals outperform normalising flow proposals.
Various possible reasons for this follow.

Firstly, fitting mixtures using the EM algorithm is relatively fast and robust compared
to fitting normalising flows using stochastic gradient optimisation.
A normalising flow can train poorly for various reasons
such as poor selection of tuning choices (including the flow architecture)
or converging to a poor local mode.
A flow proposal may perform better for models with a long simulation time,
as the time to train a proposal would then be less important.

Secondly, normalising flows may simply require larger training sets to fit all the neural network weights.
Most application of flows in machine learning use much larger training sets.
Additionally, unlike mixtures, normalising flow training sacrifices some training particles to use as a test set.

Finally, it is interesting that mixture proposals do well using only 5 mixture components,
even though this would not be enough to represent sufficiently complex multivariate distributions.
This may reflect that ABC-SMC does not reduce $\epsilon$ enough in a feasible time for the target to become this complex.
Instead the target \eqref{eq:ABC_posterior} is a smoothed approximation of the posterior,
which may be more feasible to fit by a mixture than the posterior itself.

A potential advantage of normalising flows over mixtures is that they can easily adapt to parameter constraints.
However in practice parameter transformations can easily be used to avoid issues with constraints
e.g.~transforming non-negative $\theta$ to unconstrained $\phi = \log \theta$.

\subsection{Choice of kernel family}

We argue that the one-hit approach is the most robust choice of kernel family,
as it is a competitive choice for all proposals we consider.
The ABC-MH and independence one-hit kernels also perform well for particular proposals.

We find the single proposal $r$-hit kernel can be very inefficient (Section \ref{sec:inefficiency}).
This is because waiting for accepted simulations at a poor $\theta'$ proposal can be very time consuming.

The multiple proposal $r$-hit kernel also needs to wait for accepted simulations at $\theta'$ proposals,
but makes a new proposal for each simulation.
Our experiments find this avoids the very poor performance of the single proposal version.
Nonetheless, it performs less well than other kernel families.
This suggest that the cost of requiring $r=2$ accepted simulations under $\theta'$
outweighs the potential benefits of utilising some information on the magnitude of the likelihood.

It would be interesting to further investigate the robust performance of the one-hit kernel
compared to the independence one-hit kernel family.
One possible reason is its use of early rejection (Section \ref{sec:early_rejection}).
Another is the fact it performs simulations under both $\theta$ (existing parameter value)
and $\theta'$ (proposed parameter value),
allowing relatively quick rejection when $\theta$ has a higher likelihood.
It would also be interesting to determine whether any advantage of the one-hit kernel
depends on particular features of the likelihood or proposal.
For instance, is it more robust to poor proposals?

Finally, we note that our results provide little insight into mixing.
It is unclear if and when mixing properties influence performance in our results,
especially since only one combination -- ABC-MH with classic random walk proposal --
is guaranteed to have poor mixing properties.

\subsection{Limitations}

A potential limitation of our work is that our results may not generalise beyond the models in our tests.
We have tried to mitigate this by selecting a range of models including
simple illustrations (GM, quadratic),
a more complex artificial model (SLCP)
and real applications (MG1, SEIR).

Another issue is that ABC-SMC can be less efficient than modern simulation based inference (SBI) methods
\citep{lueckmann2021benchmarking, frazier2024statistical}.
However we argue investigating its behaviour is still relevant.
Firstly it is still common in important applications such as infectious disease epidemiology \citep{ellis2024inferring}.
Secondly, the ABC posterior \eqref{eq:ABC_posterior} is of interest beyond approximate inference,
for instance in recent robust inference work \citep{miller2019robust}.

\subsection{Future work}

Here we describe several ways our work could be extended in future.

A potential further empirical investigation is how well our kernels work in ABC-MCMC.
Amongst other things, this could give some insight into their mixing properties.
Further empirical study of the low $\epsilon$ regime described in Section \ref{sec:loss_results}
could also be useful to determine what kernel features are useful here.

Also, our experiments focus on sequential computation.
SMC algorithms in general can benefit from parallel computing, by updating particles in parallel.
It would be interesting to investigate the effect of parallelisation
and which kernels benefit most.

We could consider variations on the ABC-SMC algorithm,
such as those described in Section \ref{sec:ABCSMCvariations}.
Some of the kernels described in this paper generalise naturally to these settings,
but others would need to be modified.

We currently use generic methods to train mixture proposals.
These could be modified to take the structure of the ABC-SMC algorithm into account.
For instance we could consider incrementing the number of mixture components in each SMC iteration.
This could be cheaper than searching a large range of components counts
as we investigated earlier.

We consider fitting independence proposals using Gaussian mixtures and normalising flows,
but other approaches could also be considered.
One possibility is a mixture of experts, which performs well in a SBI context \citep{haggstrom2024fast}.

It may be possible to make the normalising flow proposal more effective.
For instance perhaps the training data could be based on all previous SMC iterations,
rather than the most recent iteration as in our work.
Alternatively it may be possible to adapt the method of \cite{matthews2022continual}
which involves generating multiple sets of particles when required for training,
rather than just using the SMC particles from the previous iteration.

\paragraph{Acknowledgements}
The authors would like to thank
Pierre Jacob, Umberto Picchini and Massimiliano Tamborrino for helpful comments,
and Anthony Lee for suggesting Algorithm \ref{alg:one_hit_ind}.

Dennis Prangle and Sammy Ragy were supported by the EPSRC (grant number EP/V049127/1).

\begin{appendix}
    \section{SEIR model details} \label{app:SEIRdetails}

    Here we give more details of our SEIR epidemic model. See \citet{tang2020review} for further details.

    Recall that a closed population is divided into four compartments:
    susceptible, exposed, infectious and recovered.
    At time $t=0,1,\ldots,T$, there are $S_t$ susceptible, $E_t$ exposed, $I_t$ infectious and $R_t$ recovered individuals.
    We take $S_0=990$, $E_0=10$.
    The overall population size is fixed at $N=1000$.

    Let $\alpha^{-1}$ denote the latency period, $\beta$ and $\gamma$ the infection and recovery rates.
    At time $t$, susceptible individuals become exposed with probability $\pi_{SE}(t)= 1-\exp(-\beta I_{t-1}/N )$,
    exposed individuals become infectious with probability $\pi_{EI}=1-\exp(-\alpha)$,
    and infectious individuals recover with probability $\pi_{IR}=1-\exp(-\gamma)$.

    Observed data $y_{0:T}=y_0,\ldots,y_T$ are reported new cases.
    We assume  $Y_t\sim \text{Poisson}(\lambda_0 +\lambda\, i_t)$, where $i_t$ are new infectious cases at time $t$,
    $\lambda$ and $\lambda_0$ are the observation rate and the background rate (here $\lambda_0=0.1$ and $\lambda=0.5$).
    We infer $\theta=(\log \alpha, \log \beta, \log \gamma)$.

    Our observations are a synthetic dataset generated under parameters $\theta^{\text{true}}=(-0.5,-1,-3)$.
    Prior distributions are independent Gaussian distributions centred on $\theta^{\text{true}}$ with standard deviation equal to 2.

    \section{Tuning choices} \label{app:tuning}

    This appendix summarises tuning choices listed throughout the paper.
    It contains our default choices,
    some of which are varied in our experiments.

    \paragraph*{ABC-SMC}
    \begin{itemize}
        \item We use 1000 particles.
              For transport ABC this is split into 900 train particles and 100 test particles.
        \item We use systematic resampling.
        \item The target fraction of unique particles used when updating the threshold
              (see Section \ref{sec:threshold}) is $\omega = 0.5$.
        \item We terminate each ABC-SMC algorithm after 1 hour,
              and report results for the last complete iteration.
    \end{itemize}

    \paragraph*{Kernel families}
    \begin{itemize}
        \item We require $r=2$ simulation successes.
    \end{itemize}

    \paragraph*{Proposals}
    \begin{itemize}
        \item We use 5 components for the mixture proposal,
              following \citep{bernton2019approximate}, except when noted otherwise.
              Otherwise, we use the default choices from the AutoGMM package.
        \item Our default choice of normalising flow architecture is a rational quadratic spline flow \citep{durkan2019neural}.
              As tuning choices we use 50 bins, linear tails and a tail bound of 10.
              The neural network generating the spline has an initial fully connected layer with 20 hidden features,
              followed by 2 residual blocks.
        \item When training proposals we use $\mathcal{D}_t = \mathcal{B}_t$,
              as defined in Section \ref{sec:training_data}, except when noted otherwise.

    \end{itemize}

    \section{Training target} \label{app:target}

    Section \ref{sec:training_data} describes two possible training sets for proposal kernels
    which have been used in prior work:
    $\mathcal{A}_t$ and $\mathcal{B}_t$.
    Here we run experiments comparing them.
    We look at every model and choice of proposal kernel.
    Five replications were run for each combination, using shared random seeds as earlier.
    Throughout we use the 1-hit kernel family.

\begin{figure*}[htbp]
        \includegraphics[width=\mytextwidth]{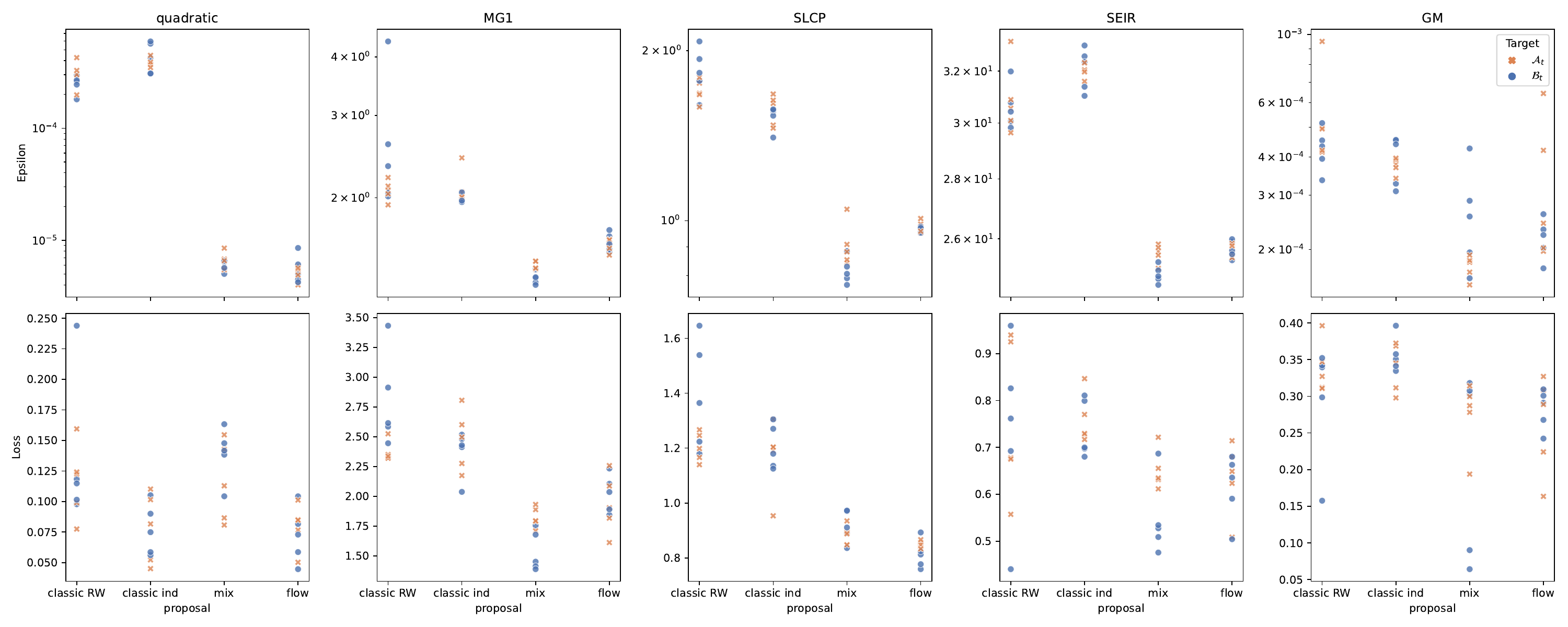}
        \caption{
            Results from varying the training target.
            Final $\epsilon$ (top, log scale) and Wasserstein loss (bottom, linear scale) values are shown.
            Each point represents a single experiment.
        } \label{fig:target_plot}
    \end{figure*}

    Figure \ref{fig:target_plot} shows the results.
    There is a lot of overlap between results for the two choices,
    and neither has a consistent advantage.

    Since we have replications with shared random seeds for each combination of model and proposal, we perform paired t-tests.
    At the 5\% level we find no significant difference in Wasserstein loss for any combination.
    For final $\epsilon$ there are two significant differences:
    \begin{itemize}
        \item MG1 with mixture proposals where $\mathcal{B}_t$ has a lower mean value ($p=0.013$)
        \item SEIR with mixture proposals where $\mathcal{B}_t$ has a lower mean value ($p=0.045$)
    \end{itemize}
    These are the only significant results from 20 tests on $\epsilon$, without a correction for multiple comparisons.
    (Using the Bonferroni adjustment, the threshold becomes $0.0025$ so neither would be significant.)

    Hence overall we interpret our experimental results as showing that the choice of training target is not critical.
    See Section \ref{sec:training_data} for more discussion on the choice of training target,
    including a justification for using $\mathcal{B}_t$ elsewhere in the paper
    based on these experimental findings.

    \section{Defensive proposal} \label{app:defensive}

    Section \ref{sec:defensive} defines a defensive proposal \eqref{eq:defensive_q}:
    a mixture of the prior and another independence proposal.
    Using it improves theoretical mixing properties of two kernels (ABC-MH and independence one-hit).
    This section investigates its empirical performance by
    repeating the experiments of Section \ref{sec:main_comparison} for a defensive version of the mixture proposal
    with a mixture weight of $\eta=0.1$ for the prior.

\begin{figure*}[htbp]
        \includegraphics[width=\mytextwidth]{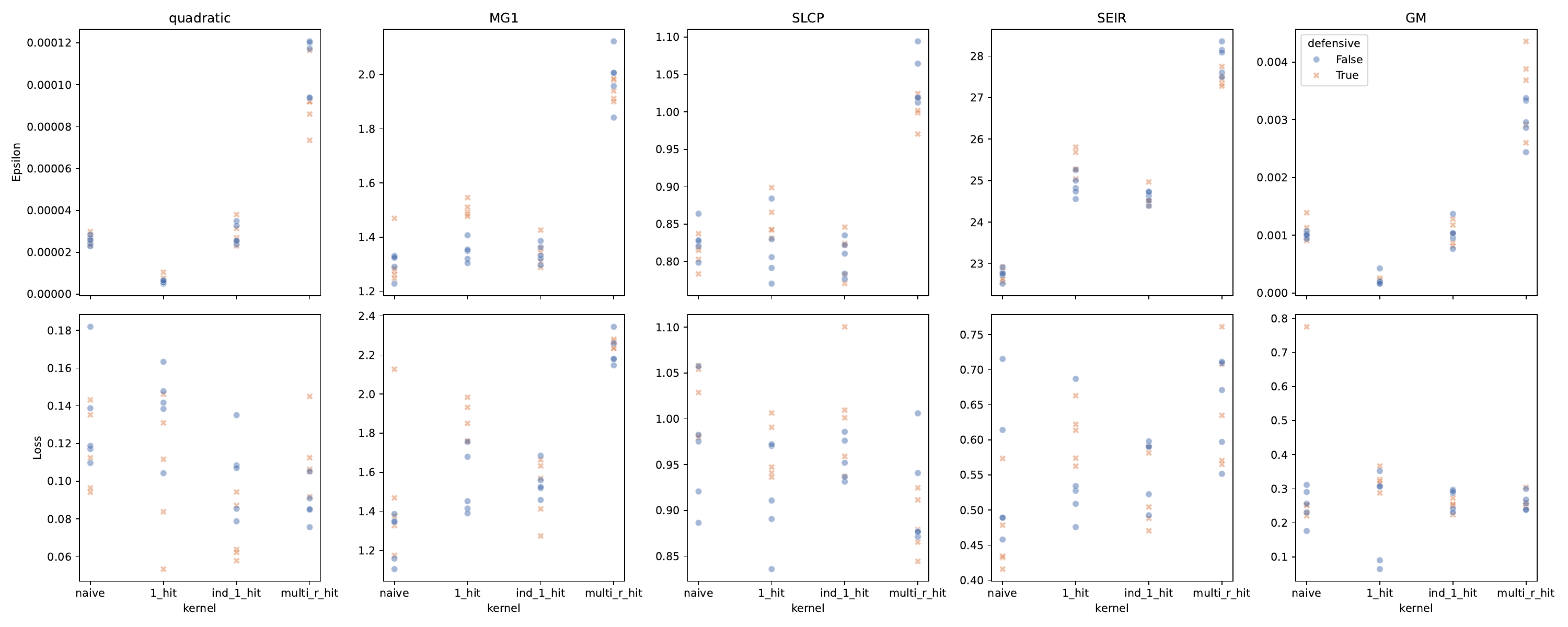}
        \caption{
            Results comparing a standard mixture proposal with a defensive variant.
            Final $\epsilon$ (top) and Wasserstein loss (bottom) values are shown.
            Each point represents a single experiment.
            There are 5 replications of each experiment.
        } \label{fig:defensive_plot}

        \includegraphics[width=\mytextwidth]{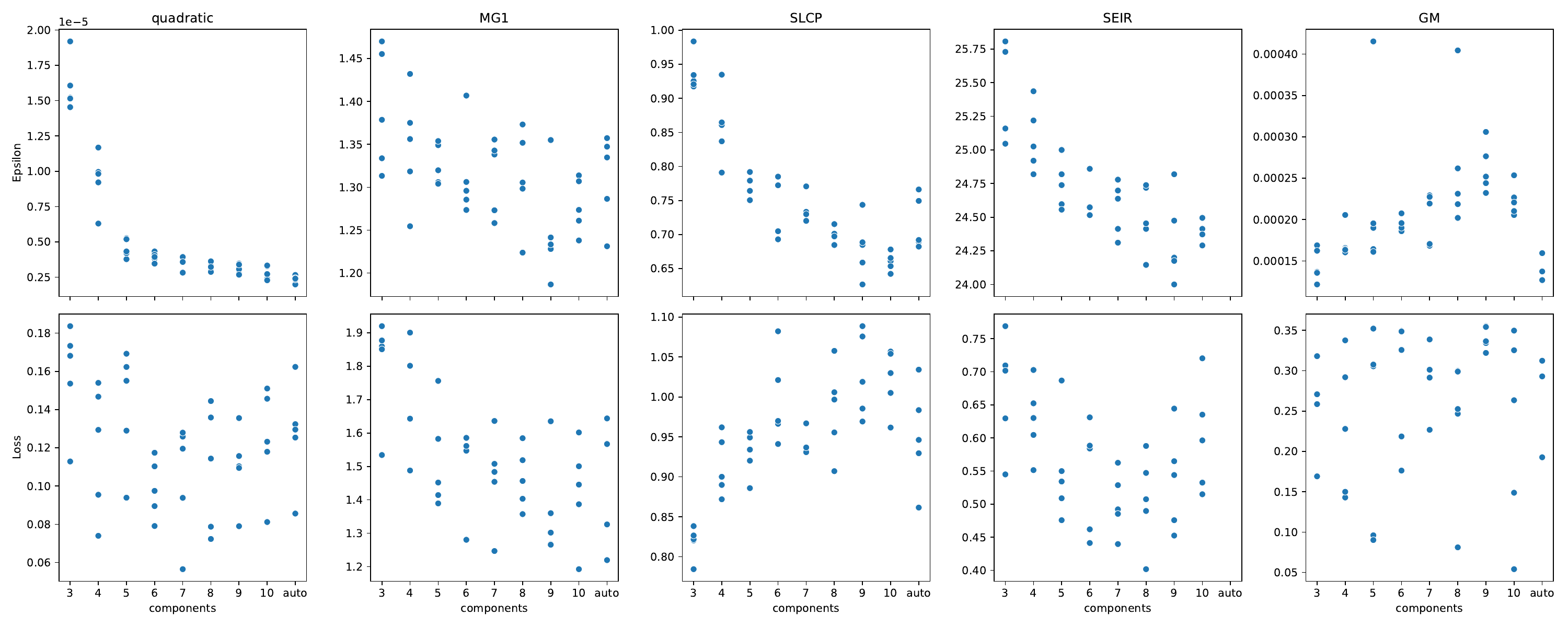}
        \caption{
            Results from varying the number of mixture components.
            Final $\epsilon$ (top) and Wasserstein loss (bottom) values are shown.
            Each point represents a single experiment.
            There are 5 replications of each experiment.
            Some points are missing for ``auto'',
            representing experiments which crashed.
        } \label{fig:components_plot}
    \end{figure*}

    Figure \ref{fig:defensive_plot} shows this produces little difference to results from the standard mixture proposal.
    To check further we ran paired $t$-tests comparing Wasserstein loss and final $\epsilon$.
    At the 5\% level we find only the following significant differences:
    \begin{itemize}
        \item MG1 with 1 hit kernel where the defensive proposal has a higher mean $\epsilon$ ($p=0.004$)
              and loss ($p=0.02$)
        \item Quadratic with multiple $r$-hit kernel where the defensive proposal has a lower mean $\epsilon$ ($p=0.02$)
    \end{itemize}
    These are the only significant results from 40 tests, without a correction for multiple comparisons.
    (Using the Bonferroni adjustment, the threshold becomes $0.00125$ so neither would be significant.)

    Overall the results show the defensive proposal results in little change in empirical performance,
    while providing better theoretical mixing guarantees for two kernels.
    However, for empirical performance alone, we find no reason to use a defensive proposal.

    \section{Underfitting and fallback} \label{app:fallback}

    Section \ref{sec:tabc} uses early stopping to avoid producing poor normalising flow proposal distributions
    by overfitting to training data.
    Another potential problem is \textbf{underfitting}.
    Here the proposal is still a poor fit to the training data once training finishes.
    This is particularly an issue for normalising flows, where training could converge to a poor local mode.

    To address this, we consider a \textbf{fallback} mechanism.
    After step 9 of Algorithm \ref{alg:TABC}, we compare the test loss of our proposal
    to the classic independence proposal (Section \ref{sec:kde}).
    We take $q_t$ to be whichever proposal produces a smaller test loss.

    Fallback should ensure the proposal is always at least as good as the classic independence proposal.
    However implementing fallback adds extra computation time.
    We experiment to test whether this is a worthwhile trade-off.

    We performed experiments for every model, using the same setup as Section \ref{sec:main_comparison}. 
    Table \ref{tab:fallback} shows the results compared to not using fallback.
    Fallback slightly improves the final $\epsilon$ in 3/5 models, but worsens it in 1/5, with 1 tie (to 4 significant figures).
    Wasserstein loss improves for 3 models and worsens for 2.
    Since we have replications with shared random seeds for each model, we perform paired t-tests.
    These showed no significant difference (at the 5\% level) in $\epsilon$ or loss value for any model.

    So overall there isn't convincing evidence of fallback improving results,
    and in the main paper we don't use it due to the extra complexity it introduces to the algorithm.
    Exploratory investigation using a version of fallback for mixture proposals also showed no clear evidence of improvement.
    A potential explanation for the lack of improvement is that underfitting is rare,
    and SMC can cope with the occasional resulting poor proposal distributions.

    \begin{table}[htbp]
        \small
        \begin{center}
            \begin{tabular}{l|rr|rr}
                \toprule
                          & \multicolumn{2}{c|}{Final $\epsilon$} & \multicolumn{2}{c}{Wasserstein loss}                           \\
                model     & No fallback                           & Fallback                             & No fallback & Fallback  \\
                \midrule
                GM        & 2.181e-04                             & 2.094e-04                            & 0.2823      & 0.2910    \\
                MG1       & 1.607                                 & 1.607                                & 2.037       & 1.916     \\
                quadratic & 6.671e-06                             & 6.389e-06                            & 7.646e-02   & 7.360e-02 \\
                SEIR      & 25.67                                 & 25.58                                & 0.6147      & 0.6396    \\
                SLCP      & 0.9810                                & 0.9915                               & 0.8165      & 0.8041    \\
                \bottomrule
            \end{tabular}
            \caption{Fallback experiment results: means of final $\epsilon$ and Wasserstein loss values to 4 significant figures
                (more precision than in other tables as many differences here are small).}
            \label{tab:fallback}
        \end{center}
    \end{table}

    \section{Further mixture component results} \label{app:extra_results}

    Figure \ref{fig:components_plot} shows further results from varying the number of mixture components,
    supplementing Section \ref{sec:mixture_components}.
\end{appendix}

\bibliographystyle{plainnat}
\bibliography{kernels}

@article{picchini2022guided,
  title={Guided sequential {ABC} schemes for intractable {Bayesian} models},
  author={Picchini, Umberto and Tamborrino, Massimiliano},
  journal={Bayesian Analysis (to appear)},
  year={2024},
  publisher={International Society for Bayesian Analysis}
}

@article{hesterberg1995weighted,
  title={Weighted average importance sampling and defensive mixture distributions},
  author={Hesterberg, Tim},
  journal={Technometrics},
  volume={37},
  number={2},
  pages={185--194},
  year={1995},
  publisher={Taylor \& Francis}
}

@inproceedings{arbel2021annealed,
  title={Annealed flow transport {Monte Carlo}},
  author={Arbel, Michael and Matthews, Alex and Doucet, Arnaud},
  booktitle={International Conference on Machine Learning},
  pages={318--330},
  year={2021},
  organization={PMLR}
}

@article{cao2024using,
author = {Cao, Xuefei and Wang, Shijia and Zhou, Yongdao},
title = {Using early rejection {Markov chain Monte Carlo} and {Gaussian} processes to accelerate {ABC} methods},
journal = {Journal of Computational and Graphical Statistics},
year = {2024},
publisher = {ASA Website},
doi = {10.1080/10618600.2024.2379349}
}

@article{del2012adaptive,
  title={An adaptive sequential {Monte Carlo} method for approximate {Bayesian} computation},
  author={Del Moral, Pierre and Doucet, Arnaud and Jasra, Ajay},
  journal={Statistics and computing},
  volume={22},
  number={5},
  pages={1009--1020},
  year={2012},
  publisher={Springer}
}

@article{beaumont2009adaptive,
  title={Adaptive approximate {Bayesian} computation},
  author={Beaumont, Mark A. and Cornuet, Jean-Marie and Marin, Jean-Michel and Robert, Christian P.},
  journal={Biometrika},
  volume={96},
  number={4},
  pages={983--990},
  year={2009},
  publisher={Oxford University Press}
}

@article{bernton2019approximate,
    author = {Bernton, Espen and Jacob, Pierre E. and Gerber, Mathieu and Robert, Christian P.},
    title = {Approximate {Bayesian} Computation with the {Wasserstein} Distance},
    journal = {Journal of the Royal Statistical Society Series B: Statistical Methodology},
    volume = {81},
    number = {2},
    pages = {235-269},
    year = {2019},
    issn = {1369-7412},
    doi = {10.1111/rssb.12312},
}

@inproceedings{lee2012choice,
  title={On the choice of {MCMC} kernels for approximate {Bayesian} computation with {SMC} samplers},
  author={Lee, Anthony},
  booktitle={Proceedings of the 2012 Winter simulation conference (WSC)},
  pages={1--12},
  year={2012},
  organization={IEEE}
}

@article{filippi2013optimality,
  title={On optimality of kernels for approximate {B}ayesian computation using sequential {Monte Carlo}},
  author={Filippi, Sarah and Barnes, Chris P. and Cornebise, Julien and Stumpf, Michael P. H.},
  journal={Statistical applications in genetics and molecular biology},
  volume={12},
  number={1},
  pages={87--107},
  year={2013},
  publisher={De Gruyter}
}

@article{toni2009approximate,
  title={Approximate {Bayesian} computation scheme for parameter inference and model selection in dynamical systems},
  author={Toni, Tina and Welch, David and Strelkowa, Natalja and Ipsen, Andreas and Stumpf, Michael P. H.},
  journal={Journal of the Royal Society Interface},
  volume={6},
  number={31},
  pages={187--202},
  year={2009},
  publisher={The Royal Society London}
}

@article{sisson2007sequential,
  title={Sequential {Monte Carlo} without likelihoods},
  author={Sisson, S. A. and Fan, Y. and Tanaka, Mark M.},
  journal={Proceedings of the National Academy of Sciences},
  volume={104},
  number={6},
  pages={1760--1765},
  year={2007},
  publisher={National Acad Sciences}
}

@article{del2006sequential,
  title={Sequential {Monte Carlo} samplers},
  author={Del Moral, Pierre and Doucet, Arnaud and Jasra, Ajay},
  journal={Journal of the Royal Statistical Society Series B: Statistical Methodology},
  volume={68},
  number={3},
  pages={411--436},
  year={2006},
  publisher={Oxford University Press}
}

@article{beskos2016on,
author = {Alexandros Beskos and Ajay Jasra and Nikolas Kantas and Alexandre Thiery},
title = {On the convergence of adaptive sequential {Monte Carlo} methods},
volume = {26},
journal = {The Annals of Applied Probability},
number = {2},
publisher = {Institute of Mathematical Statistics},
pages = {1111--1146},
year = {2016},
}

@article{everitt2021delayed,
author = {Richard G. Everitt and Paulina A. Rowińska},
title = {Delayed Acceptance {ABC-SMC}},
journal = {Journal of Computational and Graphical Statistics},
volume = {30},
number = {1},
pages = {55-66},
year  = {2021},
publisher = {Taylor & Francis}
}

@article{carpenter1999improved,
  title={Improved particle filter for nonlinear problems},
  author={Carpenter, James and Clifford, Peter and Fearnhead, Paul},
  journal={IEE Proceedings-Radar, Sonar and Navigation},
  volume={146},
  number={1},
  pages={2--7},
  year={1999},
  publisher={IET}
}

@inproceedings{matthews2022continual,
  title={Continual repeated annealed flow transport {Monte Carlo}},
  author={Matthews, Alex and Arbel, Michael and Rezende, Danilo Jimenez and Doucet, Arnaud},
  booktitle={International Conference on Machine Learning},
  pages={15196--15219},
  year={2022},
  organization={PMLR}
}

@incollection{sisson2011likelihood,
	doi = {10.1201/b10905-13},
	year = 2011,
	pages = {313--335},
	publisher = {Chapman and Hall/{CRC}},
	author = {Scott Sisson and Yanan Fan},
	title = {Likelihood-Free {MCMC}},
	booktitle = {Handbook of {Markov} Chain {Monte Carlo}}
}

@article{marjoram2003markov,
	doi = {10.1073/pnas.0306899100},
	year = 2003,
	publisher = {Proceedings of the National Academy of Sciences},
	volume = {100},
	number = {26},
	pages = {15324--15328},
	author = {Paul Marjoram and John Molitor and Vincent Plagnol and Simon Tavar{\'{e}}},
	title = {Markov chain {Monte Carlo} without likelihoods},
	journal = {Proceedings of the National Academy of Sciences}
}

@article{marin2012approximate,
	doi = {10.1007/s11222-011-9288-2},
	year = 2012,
	publisher = {Springer Science and Business Media {LLC}},
	volume = {22},
	number = {6},
	pages = {1167--1180},
	author = {Jean-Michel Marin and Pierre Pudlo and Christian P. Robert and Robin J. Ryder},
	title = {Approximate {Bayesian} computational methods},
	journal = {Statistics and Computing}
}

@incollection{sisson2018abc,
	doi = {10.1201/9781315117195-4},
	year = 2018,
	publisher = {Chapman and Hall/{CRC}},
	pages = {87--123},
	author = {S. A. Sisson and Y. Fan},
	title = {{ABC} Samplers},
	booktitle = {Handbook of Approximate Bayesian Computation}
}

@article{prangle2017adapting,
	doi = {10.1214/16-ba1002},
	year = 2017,
	publisher = {Institute of Mathematical Statistics},
	volume = {12},
	number = {1},
	author = {Dennis Prangle},
	title = {Adapting the {ABC} Distance Function},
	journal = {Bayesian Analysis}
}

@book{chopin2020introduction,
	doi = {10.1007/978-3-030-47845-2},
	year = 2020,
	publisher = {Springer International Publishing},
	author = {Nicolas Chopin and Omiros Papaspiliopoulos},
	title = {An Introduction to {Sequential Monte Carlo}}
}

@article{gerber2019negative,
	doi = {10.1214/18-aos1746},
	year = 2019,
	publisher = {Institute of Mathematical Statistics},
	volume = {47},
	number = {4},
	author = {Mathieu Gerber and Nicolas Chopin and Nick Whiteley},
	title = {Negative association, ordering and convergence of resampling methods},
	journal = {The Annals of Statistics}
}

@article{picchini2014inference,
author = {Umberto Picchini},
title = {Inference for SDE Models via Approximate Bayesian Computation},
journal = {Journal of Computational and Graphical Statistics},
volume = {23},
number = {4},
pages = {1080--1100},
year = {2014},
publisher = {ASA Website},
doi = {10.1080/10618600.2013.866048}
}

@article{banterle2015accelerating,
  title={Accelerating {Metropolis-Hastings} algorithms by delayed acceptance},
  author={Banterle, Marco and Grazian, Clara and Lee, Anthony and Robert, Christian P.},
  journal={arXiv preprint arXiv:1503.00996},
  year={2015}
}

@Book{scrucca2023model,
  title = {Model-Based Clustering, Classification, and Density Estimation Using {mclust} in {R}},
  author = {Luca Scrucca and Chris Fraley and T. Brendan Murphy and Adrian E. Raftery},
  publisher = {Chapman and Hall/CRC},
  isbn = {978-1032234953},
  doi = {10.1201/9781003277965},
  year = {2023},
  url = {https://mclust-org.github.io/book/},
}

@article{athey2019autogmm,
  title={{AutoGMM}: Automatic and hierarchical {Gaussian} mixture modeling in {Python}},
  author={Athey, Thomas L. and Liu, Tingshan and Pedigo, Benjamin D. and Vogelstein, Joshua T.},
  journal={arXiv preprint arXiv:1909.02688},
  year={2019}
}

@article{kobyzev2020normalizing,
  title={Normalizing flows: An introduction and review of current methods},
  author={Kobyzev, Ivan and Prince, Simon J. D. and Brubaker, Marcus A.},
  journal={IEEE transactions on pattern analysis and machine intelligence},
  volume={43},
  number={11},
  pages={3964--3979},
  year={2020},
  publisher={IEEE}
}

@inproceedings{lueckmann2021benchmarking,
  title={Benchmarking Simulation-Based Inference},
  author={Jan-Matthis Lueckmann and Jan Boelts and David S. Greenberg and Pedro J. Gonçalves and Jakob H. Macke},
  booktitle={International Conference on Artificial Intelligence and Statistics},
  year={2021},
  url={https://api.semanticscholar.org/CorpusID:231583170}
}

@article{papamakarios2021normalizing,
  title={Normalizing flows for probabilistic modeling and inference},
  author={Papamakarios, George and Nalisnick, Eric and Rezende, Danilo Jimenez and Mohamed, Shakir and Lakshminarayanan, Balaji},
  journal={The Journal of Machine Learning Research},
  volume={22},
  number={1},
  pages={2617--2680},
  year={2021},
  publisher={JMLRORG}
}

@inproceedings{papamakarios2019sequential,
  title={Sequential neural likelihood: Fast likelihood-free inference with autoregressive flows},
  author={Papamakarios, George and Sterratt, David and Murray, Iain},
  booktitle={The 22nd international conference on artificial intelligence and statistics},
  pages={837--848},
  year={2019},
  organization={PMLR}
}

@article{papamakarios2016fast,
  title={Fast $\varepsilon$-free inference of simulation models with {Bayesian} conditional density estimation},
  author={Papamakarios, George and Murray, Iain},
  journal={Advances in neural information processing systems},
  volume={29},
  year={2016}
}

@incollection{prangle2018summary,
  title={Summary statistics},
  author={Prangle, Dennis},
  booktitle={Handbook of approximate Bayesian computation},
  pages={125--152},
  year={2018},
  publisher={Chapman and Hall/CRC}
}

@inproceedings{chen2023learning,
  title={Is learning summary statistics necessary for likelihood-free inference?},
  author={Chen, Yanzhi and Gutmann, Michael U. and Weller, Adrian},
  booktitle={International Conference on Machine Learning},
  pages={4529--4544},
  year={2023},
  organization={PMLR}
}

@article{fearnhead2012constructing,
  title={Constructing summary statistics for approximate Bayesian computation: semi-automatic approximate Bayesian computation},
  author={Fearnhead, Paul and Prangle, Dennis},
  journal={Journal of the Royal Statistical Society Series B: Statistical Methodology},
  volume={74},
  number={3},
  pages={419--474},
  year={2012},
  publisher={Oxford University Press}
}

@article{durkan2019neural,
  title={Neural spline flows},
  author={Durkan, Conor and Bekasov, Artur and Murray, Iain and Papamakarios, George},
  journal={Advances in neural information processing systems},
  volume={32},
  year={2019}
}

@article{prangle2023distilling,
  title={Distilling importance sampling for likelihood free inference},
  author={Prangle, Dennis and Viscardi, Cecilia},
  journal={Journal of Computational and Graphical Statistics},
  volume={32},
  number={4},
  pages={1461--1471},
  year={2023},
  publisher={Taylor \& Francis}
}

@article{shestopaloff2014bayesian,
  title={On {Bayesian} inference for the {M/G/1} queue with efficient {MCMC} sampling},
  author={Shestopaloff, Alexander Y. and Neal, Radford M.},
  journal={arXiv preprint arXiv:1401.5548},
  year={2014}
}

@article{andrieu2010particle,
  title={Particle {Markov} chain {Monte Carlo} methods},
  author={Andrieu, Christophe and Doucet, Arnaud and Holenstein, Roman},
  journal={Journal of the Royal Statistical Society Series B: Statistical Methodology},
  volume={72},
  number={3},
  pages={269--342},
  year={2010},
  publisher={Oxford University Press}
}

@article{beaumont2019approximate,
  title={Approximate {Bayesian} computation},
  author={Beaumont, Mark A.},
  journal={Annual review of statistics and its application},
  volume={6},
  number={1},
  pages={379--403},
  year={2019},
  publisher={Annual Reviews}
}

@article{drovandi2022comparison,
  title={A comparison of likelihood-free methods with and without summary statistics},
  author={Drovandi, Christopher and Frazier, David T.},
  journal={Statistics and Computing},
  volume={32},
  number={3},
  pages={42},
  year={2022},
  publisher={Springer}
}

@article{lee2014variance,
  title={Variance bounding and geometric ergodicity of {Markov chain Monte Carlo} kernels for approximate {Bayesian} computation},
  author={Lee, Anthony and {\L}atuszy{\'n}ski, Krzysztof},
  journal={Biometrika},
  volume={101},
  number={3},
  pages={655--671},
  year={2014},
  publisher={Oxford University Press}
}

@InProceedings{kingma2015adam,
  author    = {Kingma, Diederik and Ba, Jimmy},
  booktitle = {International Conference on Learning Representations (ICLR)},
  title     = {Adam: A Method for Stochastic Optimization},
  year      = {2015},
  address   = {San Diega, CA, USA},
}

@article{flamary2021pot,
  author  = {R{\'e}mi Flamary and Nicolas Courty and Alexandre Gramfort and Mokhtar Z. Alaya and Aur{\'e}lie Boisbunon and Stanislas Chambon and Laetitia Chapel and Adrien Corenflos and Kilian Fatras and Nemo Fournier and L{\'e}o Gautheron and Nathalie T.H. Gayraud and Hicham Janati and Alain Rakotomamonjy and Ievgen Redko and Antoine Rolet and Antony Schutz and Vivien Seguy and Danica J. Sutherland and Romain Tavenard and Alexander Tong and Titouan Vayer},
  title   = {POT: Python Optimal Transport},
  journal = {Journal of Machine Learning Research},
  year    = {2021},
  volume  = {22},
  number  = {78},
  pages   = {1-8},
  url     = {http://jmlr.org/papers/v22/20-451.html}
}

@article{mengersen1996rates,
  title={Rates of convergence of the {Hastings} and {Metropolis} algorithms},
  author={Mengersen, Kerrie L. and Tweedie, Richard L.},
  journal={The Annals of Statistics},
  volume={24},
  number={1},
  pages={101--121},
  year={1996},
  publisher={Institute of Mathematical Statistics}
}

@article{whiteley2012sequential,
  title={Sequential {Monte Carlo} samplers: error bounds and insensitivity to initial conditions},
  author={Whiteley, Nick},
  journal={Stochastic Analysis and Applications},
  volume={30},
  number={5},
  pages={774--798},
  year={2012},
  publisher={Taylor \& Francis}
}

@article{dau2022waste,
  title={Waste-free sequential {Monte Carlo}},
  author={Dau, Hai-Dang and Chopin, Nicolas},
  journal={Journal of the Royal Statistical Society Series B: Statistical Methodology},
  volume={84},
  number={1},
  pages={114--148},
  year={2022},
  publisher={Oxford University Press}
}

@article{beskos2014stability,
  title={On the stability of sequential {Monte Carlo} methods in high dimensions},
  volume={24},
  number={4},
  journal={The Annals of Applied Probability},
  publisher={Institute of Mathematical Statistics},
  author={Beskos, Alexandros and Crisan, Dan and Jasra, Ajay},
  year={2014},
}

@article{miller2019robust,
  title={Robust {Bayesian} inference via coarsening},
  author={Miller, Jeffrey W and Dunson, David B.},
  journal={Journal of the American Statistical Association},
  year={2019},
  publisher={Taylor \& Francis}
}

@article{frazier2024statistical,
  title={The Statistical Accuracy of Neural Posterior and Likelihood Estimation},
  author={Frazier, David T. and Kelly, Ryan and Drovandi, Christopher and Warne, David J.},
  journal={arXiv preprint arXiv:2411.12068},
  year={2024}
}

@article{ellis2024inferring,
  title={Inferring transmission routes for foot-and-mouth disease virus within a cattle herd using approximate {Bayesian} computation},
  author={Ellis, John and Brown, Emma and Colenutt, Claire and Schley, David and Gubbins, Simon},
  journal={Epidemics},
  volume={46},
  pages={100740},
  year={2024},
  publisher={Elsevier}
}

@article{haggstrom2024fast,
  title={Fast, accurate and lightweight sequential simulation-based inference using {Gaussian} locally linear mappings},
  author={H{\"a}ggstr{\"o}m, Henrik and Rodrigues, Pedro LC and Oudoumanessah, Geoffroy and Forbes, Florence and Picchini, Umberto},
  journal={Transactions on Machine Learning Research},
  issn={2835-8856},
  year={2024}
}

@article{wilkinson2013approximate,
  title={Approximate {Bayesian} computation ({ABC}) gives exact results under the assumption of model error},
  author={Wilkinson, Richard David},
  journal={Statistical applications in genetics and molecular biology},
  volume={12},
  number={2},
  pages={129--141},
  year={2013},
  publisher={De Gruyter}
}

@book{villani2009optimal,
  title={Optimal transport: old and new},
  author={Villani, C{\'e}dric},
  year={2008},
  publisher={Springer Berlin Heidelberg}
}

@article{tang2020review,
author = {Tang, Lu and Zhou, Yiwang and Wang, Lili and Purkayastha, Soumik and Zhang, Leyao and He, Jie and Wang, Fei and Song, Peter X.-K.},
title = {A Review of Multi-Compartment Infectious Disease Models},
journal = {International Statistical Review},
volume = {88},
number = {2},
pages = {462-513},
doi = {https://doi.org/10.1111/insr.12402},
url = {https://onlinelibrary.wiley.com/doi/abs/10.1111/insr.12402},
eprint = {https://onlinelibrary.wiley.com/doi/pdf/10.1111/insr.12402},
year = {2020}
}

\end{document}